\documentclass[letterpaper,12pt,oneside]{memoir}
\usepackage[utf8x]{inputenc}
\usepackage{graphicx}
\usepackage{amsmath}
\usepackage{color}
\usepackage{amsthm}
\usepackage[caption=false,position=bottom]{subfig}
\newcommand{\myauthor}{David T. Johnson}
\newcommand{\mytitle}{Generalized Galilean Transformations and the Measurement Problem in the Entropic Dynamics Approach to Quantum Theory}
\usepackage[pdfborder={0 0 0},colorlinks=true,urlcolor=black,citecolor=black,linkcolor=black,pdftitle={\myauthor\ - \mytitle},pdfauthor={\myauthor},pdfcreator={},pdfproducer={}]{hyperref}

\newcommand{\singlespacing}{\linespread{1}}
\newcommand{\doublespacing}{%
	\linespread{2}%
	\providecommand{\doublespacingon}{1}%
}

\usepackage[left=1in,right=1in,top=1in,bottom=1in,nohead]{geometry}

\chapterstyle{thatcher}
\setsecheadstyle{\bfseries\slshape}
\setsubsecheadstyle{\itshape}

\newtheorem*{thm*}{Theorem}

\allowdisplaybreaks

\newcommand{\tsum}{\textstyle\sum}

\newcommand{\eqdef}{\overset{\underset{\mathrm{def}}{}}{=}}

\let\oldepigraph\epigraph
\renewcommand{\epigraph}[2]{{\singlespacing\oldepigraph{\textit{#1}}{-- \textsc{#2}}}}
\newcommand{\mychapter}[1]{%
	\singlespacing\chapter{#1}%
	\ifx\doublespacingon\undefined%
	\else%
	\doublespacing%
	\fi%
}

\begin{document}

\begin{titlingpage}
	\null\vfill
	\begin{center}\MakeUppercase{\large \mytitle}\end{center}
	\vfill
	\begin{center}by\end{center}
	\vfill
	\begin{center}\myauthor\end{center}
	\vfill
	\begin{center}
		A Dissertation\\
		Submitted to the University at Albany, State University of New York\\
		in Partial Fulfillment of\\
		the Requirements for the Degree of\\
		Doctor of Philosophy
	\end{center}
	\vfill
	\begin{center}
		College of Arts \& Sciences\\
		Department of Physics\\
		\the\year
	\end{center}
	\vfill\null

\end{titlingpage}

\frontmatter
\setcounter{page}{2} 

\chapter*{Abstract} 
	Quantum mechanics is an extremely successful and accurate physical theory, yet since its inception, it has been afflicted with numerous conceptual difficulties. The primary subject of this thesis is the theory of entropic quantum dynamics (EQD), which seeks to avoid these conceptual problems by interpreting quantum theory from an informational perspective.

We begin by reviewing Cox's work in describing probability theory as a means of rationally and consistently quantifying uncertainties. We then discuss how probabilities can be updated according to either Bayes' theorem or the extended method of maximum entropy (ME). After that discussion, we review the work of Caticha and Giffin that shows that Bayes' theorem is a special case of ME. This important result demonstrates that the ME method is the general method for updating probabilities.

We then review some motivating difficulties in quantum mechanics before discussing Caticha's work in deriving quantum theory from the approach of entropic dynamics, which concludes our review.

After entropic dynamics is introduced, we develop the concepts of symmetries and transformations from an informational perspective. The primary result is the formulation of a symmetry condition that any transformation must satisfy in order to qualify as a symmetry in EQD. We then proceed to apply this condition to the extended Galilean transformation. This transformation is of interest as it exhibits features of both special and general relativity. The transformation yields a gravitational potential that arises from an equivalence of information.

We conclude the thesis with a discussion of the measurement problem in quantum mechanics. We discuss the difficulties that arise in the standard quantum mechanical approach to measurement before developing our theory of entropic measurement. In entropic dynamics, position is the only observable. We show how a theory built on this one observable can account for the multitude of measurements present in quantum theory. Furthermore, we show that the Born rule need not be postulated, but can be derived in EQD. Finally, we show how the wave function can be updated by the ME method as the phase is constructed purely in terms of probabilities.
	
\chapter*{Dedication}
\begin{center}
	\textit{To Ellen}
\end{center}

\chapter*{Acknowledgments}
	I am deeply indebted to the faculty and staff at the University at Albany for their endless support and encouragement over the years.

I wish to thank all of my research committee members for their patience, flexibility, and generosity. I would particularly like to thank Professor Kevin Knuth and Professor Keith Earle for their guidance and friendship during my time at Albany. I am very grateful to Dr. Adom Giffin for graciously joining my committee and for providing invaluable advice. I would also like to thank Professor Akira Inomata for his dedication and for his continued interest in my research.

I also wish to thank Professor Carolyn MacDonald for the numerous opportunities she has offered to me and for the trust she has placed in me as my teaching supervisor and as department chair.

I cannot express how grateful I am to my advisor, Professor Ariel Caticha. His selfless dedication to his students and his teaching has an indelible impact on anyone he meets. His friendship, encouragement, and advice has helped make me into the physicist that I am today. This thesis is the result of the countless hours I spent working with Professor Caticha.

Finally, I wish to thank my friends and my family -- particularly my wife, Ellen. Without their unfailing support over the many years of my education, I would have never made it to this point.

\newpage
\tableofcontents
\newpage
\listoffigures

\mainmatter
\setcounter{secnumdepth}{2}

\mychapter{Introduction}
	\label{chapter:introduction}
	\epigraph{If your experiment needs statistics, you ought to have done a better experiment.}{Ernest Rutherford \cite{bailey:1967}}

\noindent Before the turn of the twentieth century and the advent of quantum mechanics, the use of statistics and probability in physics was regarded as an unfortunate consequence of imprecise experiments. Even the probabilistic nature of statistical mechanics could be disposed of provided one could measure the position and momentum of each particle. When quantum mechanics arrived on the scene, however, many concluded that probability must be an inherent feature of reality. Unfortunately, the great misunderstanding of probability theory polluted an otherwise quantitatively successful theory. Quantum mechanics, perhaps our best physical theory, has been plagued by countless conceptual difficulties. The result has been the formulation of numerous alternative quantum theories and interpretations---none gaining sufficient acceptance to dismiss with the alternatives.

The subject of this thesis is yet another reinterpretation of quantum theory called entropic quantum dynamics (EQD). This theory, however, is of an unprecedented nature. Entropic dynamics asserts that quantum theory is an informational theory. It arises when we attempt to make inferences with incomplete information. (Whether the missing information is even attainable is not relevant at this point.) This is the origin of the probabilistic nature of quantum mechanics.

There have been previous attempts to describe quantum mechanics as a theory of information. One such example is Ballentine's statistical interpretation of quantum mechanics \cite{ballentine:1970}. Ballentine asserts that the wave function is not physical; it only gives the probability that particles have certain properties. He also shows how many features thought to be crucial to quantum theory are unnecessary or of limited applicability \cite{ballentine:1990}. However, Ballentine subscribes to a frequentist interpretation of probability. Wave functions do not apply to individual systems but to large ensembles of similarly prepared systems. As such, many of the powerful inferential tools from the Bayesian perspective are not available. This leads Ballentine to assert postulates similar in nature to the postulates of standard quantum mechanics \cite{ballentine:1970,ballentine:1998}. However, his goal was not to derive quantum theory from more fundamental principles but rather to properly interpret the \emph{meaning} of the theory.

Entropic dynamics is also formally very similar to an alternate theory of quantum mechanics developed by Nelson \cite{nelson:1966}. Nelson's \emph{stochastic mechanics} attempts to ascribe the probabilistic nature of quantum mechanics to an underlying classical Brownian motion. While there are mathematical similarities between entropic dynamics and stochastic mechanics, the theories are very different. EQD operates at the level of information. Stochastic mechanics presumes to describe reality itself. As a result, stochastic mechanics was faced with numerous conceptual and even experimental difficulties, which ultimately led Nelson to abandon the theory. We will discuss one such issue later in chapter \ref{chapter:measurement}.

The goal of entropic quantum dynamics is not to replace quantum mechanics with an entirely alternate theory. Our goals are closer to Ballentine in that we seek to understand quantum theory in terms of information. However, we can go much further. We will show in the course of this thesis how every postulate in quantum mechanics can either be eliminated as unnecessary or be replaced by fundamental, reasonable, and informational assumptions.

This thesis is organized as follows. In chapter \ref{chapter:quantifying_uncertainty} we review Cox's work and describe probability theory as \emph{the} means of rationally and consistently assigning the plausibilities of assertions \cite{cox:1946}. This objective Bayesian approach to probability allows far more flexibility than a frequentist approach yet rejects the subjectivity present in some views of Bayesian statistics. The chapter concludes with a review of important tools in probability theory that will be used extensively in later chapters. (Much of this chapter and the next read as a `history as it should have been.' We summarize the important results and note the key minds behind them. However, we do not include the missteps nor the countless contributions from other less prominent figures that led to these results. This omission is, unfortunately, the consequence of progress.)

Chapter \ref{chapter:updating_probs} explores how one updates probabilities in an objective way. If probabilities represent one's state of knowledge, then when information is presented, the probabilities must be updated. The tool to update when information comes in the form of data is Bayes' theorem. When information comes as constraints, the tool is the extended method of maximum entropy. We briefly review Caticha's derivation of this method \cite{caticha:2003,caticha:2007}. Later in the chapter, we review the remarkable work by Giffin and Caticha that shows that Bayes' theorem is actually a special case of the maximum entropy method \cite{caticha:2006,giffin:2007}. This important discovery implies that there is only one method for updating probabilities---the method of maximum entropy. 

An additional important result in chapter \ref{chapter:updating_probs} is a review of Jaynes' treatment of statistical mechanics as an inference problem \cite{jaynes:1957}. This discovery paved the way for other theories to be cast in an epistemological light, such as our treatment of quantum mechanics as entropic dynamics.

In chapter \ref{chapter:eqd} we review Caticha's work in developing entropic quantum dynamics \cite{caticha:2009,caticha:2010}. We also discuss some of the fundamental conceptual issues in the standard quantum approach that motivate the search for an alternative theory.

In chapter \ref{chapter:symmetry} we begin the discussion of our work. The subject of this chapter is the concept of symmetry. We discuss what a symmetry means in informational terms, and formulate a symmetry condition that any transformation must obey in order to qualify as a symmetry.

In chapter \ref{chapter:galilean} we apply our symmetry condition to the extended Galilean transformation. This transformation is interesting as it admits residual effects of special and general relativity. While the behavior of the extended Galilean transformation is known in standard QM, we consider the transformation from a very fundamental point of view. The resulting equivalence between uniform gravitational fields and constantly accelerating frames implies an equivalence of information.

Finally, in chapter \ref{chapter:measurement} we explore the problem of measurement in quantum mechanics from the perspective of EQD. More than any other difficulty in quantum theory, ``the measurement problem'' has motivated numerous alternative theories. The chapter begins with some mathematical formalism before moving on to a discussion of the measurement problem in standard quantum mechanics. We then explain how measurement is handled in our entropic approach.

In entropic dynamics, position is the only observable. We will show how a theory built only on position can account for the vast array of measurements one can perform on a quantum system. In the standard quantum theory, the Born rule is a postulate that determines the probabilities of outcomes of a measurement. In entropic dynamics, however, this rule need not be postulated. For position, the rule is a direct consequence of the statistical model underlying EQD. For measurements of other observables, we derive the Born rule from the unitary evolution of the Schr\"odinger equation.

Another postulate that we examine is the projection postulate. The postulate states that after interacting with a measuring device the wave function must be left in an eigenstate of the operator representing the device. We discuss how this postulate originates when one forces a realistic interpretation on the wave function. It is reinforced by over-application of a very specialized experimental procedure known as filtering. We show how in these special cases the ME method can be used to update the wave function when new, relevant information is available. Such updating is only possible in our entropic approach because the entire wave function (including the phase) is statistical in nature.

In the final chapter, we review a representative list of the postulates underlying standard quantum mechanics. We examine each one and show how our entropic approach to quantum theory renders them unnecessary or simply consequences of more fundamental informational assumptions.

\mychapter{Quantifying Plausibilities}
	\label{chapter:quantifying_uncertainty}
	Historically, the interpretation of probability theory divides into two schools of thought \cite{caticha:2008}. The \emph{frequentist} approach views probabilities as the frequency of occurrence of random events in an infinite ensemble of sufficiently identical trials. The appeal of this interpretation is apparent when approaching problems that approximate these conditions---repeatedly tossing a coin or examining the properties of a collection of atoms, for example.

Objections to the frequentist interpretation are immediately raised. How identical is sufficiently identical? If the trials are completely identical, shouldn't the outcomes always be the same? How large of an an ensemble is large enough? A more serious deficit exists; the frequentist approach does not account for everyday usage of probabilities. Consider the question, what is the probability that it will rain today? How does one construct an ensemble in this problem? All days are clearly not identical.

The alternative view of probability is the \emph{Bayesian} interpretation. From this perspective, probabilities are measures of confidence or plausibility that an assertion is true. The name derives from an important theorem called Bayes' theorem, which will be discussed later in section \ref{sec:up_bayes}. The Bayesian interpretation is significantly more general than the frequentist approach---perhaps too general. While interpreting probabilities as plausibilities in assertions has much greater applicability, it introduces a spectrum of subjectivity. On one end of the spectrum, probabilities are viewed in a personalistic way where each individual may assign different plausibilities to the same assertion based on their own views. At the other end of the spectrum is the objective Bayesian viewpoint, which we subscribe to. This interpretation seeks to remove as much subjectivity as possible so that two individuals faced with the same information will agree on the assignment of probabilities.

The goal of this chapter is to derive a means of consistently and rationally quantifying the plausibility of statements representing a state of knowledge. We will not invoke probability theory, but the remarkable result is that the method for rationally quantifying the plausibility of assertions turns out to be the very same rules for probabilities.

\section{Eliminative Induction}
	\label{sec:quant_rational_plaus}
	
	An extremely useful tool that we will apply repeatedly is John Skilling's eliminative induction \cite{skilling:1988}. Simply stated, if a general theory exists, it should apply to special cases. This implies that one can start with a sufficiently large set of theories and by requiring that they satisfy a sufficient number of special cases, the general theory can be fully constrained by discarding those theories that are incompatible with the special cases. While this method is powerful and remarkably intuitive, it is not guaranteed to work. If there are too many special cases, incompatible special cases or if the general theory is not sufficiently general, the method will fail to capture \emph{any} theories at all.

\section{Notation}

	Given some statement $a$, we seek a function $\mathcal{P}(a)$ that assigns a real number as the plausibility of the statement in a consistent way. A useful concept is that of a conditional plausibility---the plausibility that a statement $a$ is true given that some other statement $b$ is known to be true. We write this as $\mathcal{P}(a|b)$, which is read `the plausibility of $a$ given $b$.' It should be noted that all plausibilities (and probabilities) are conditional on \emph{something}. Many texts write this as $\mathcal{P}(a|I)$, where $I$ represents all relevant background information that is known (e.g. \cite{sivia:2006}). Where there is no risk of confusion, we will suppress the use of plausibilities conditional on background information to simplify our notation.
	
	The \emph{conjunction} of two statements $a$ and $b$ is written $ab$ or $a,b$ and is true only if both $a$ and $b$ are simultaneously true and is false otherwise. The plausibility of the conjunction is then $\mathcal{P}(ab)$, read `the plausibility of $a$ \emph{and} $b$.' The plausibility of the conjunction is often called the `joint plausibility.' The \emph{disjunction} of two statements $a$ and $b$ is true if either $a$ or $b$ is true and is false only if both $a$ and $b$ are false. The disjunction is written $a+b$, and the plausibility $\mathcal{P}(a+b)$ is read `the plausibility of $a$ \emph{or} $b$.' This choice of notation for conjunctions and disjunctions will become clear when we uncover the sum and product rules later in this chapter.

	For every statement $a$ that can be true, there exists the negation `not-$a$' that must be false when $a$ is true and vice versa. We denote the negation not-$a$ as $\neg a$. The negation of the conjunction $ab$ is $\neg (ab)$, which is true when either $a$ or $b$ is false, $\neg(ab) = \neg a + \neg b$. The negation of the disjunction of $a$ and $b$ is $\neg(a+b)$, which is true only when $a$ and $b$ are both false,
	\begin{equation}
		\label{eq:disjunction_negation}
		\neg(a+b) = \neg a \neg b \ .
	\end{equation}

\section{Cox's Axioms}
	
	We wish to constrain our general plausibility function $\mathcal{P}$ by introducing rules for consistent assignment and examining special cases. These rules were originally developed by Richard Cox in 1946 as two axioms \cite{cox:1946}. The axioms simply describe how an assignment of plausibilities must behave in order to be consistent and rational.
	
	\newtheorem{cox}{Axiom}

	First, the degree of plausibility of a statement $a$ is not independent of the degree of plausibility of its negation $\neg a$. If $a$ becomes more plausible, then $\neg a$ must become less plausible, which leads us to the first axiom:
	\begin{cox}
		The plausibility of not-$a$, or $\neg a$, is a monotonic function of the plausibility of $a$,
		\begin{equation}
			\mathcal{P}(\neg a) = f(\mathcal{P}(a)) \ .
		\end{equation}
	\end{cox}
	\noindent We are not saying anything about the amount in which the plausibility of $\neg a$ must change when $a$ changes, only that they must be related by some unknown monotonic function $f$.

	The second axiom examines the plausibility that two different (but not necessarily independent) statements $a$ and $b$ are both simultaneously true. In order for the conjunction $ab$ to be true, $a$ must be true. Furthermore, once $a$ is known to be true, the statement $b$ must be true given that $a$ is true. So the plausibility of the conjunction must depend on $\mathcal{P}(a)$ and $\mathcal{P}(b|a)$, which leads to Cox's second axiom:
	\begin{cox}
		The plausibility of $ab$ is a function of the plausibility of $a$ and the plausibility of $b$ given that $a$ is true,
		\begin{equation}
			\mathcal{P}(ab) = g[\mathcal{P}(a),\mathcal{P}(b|a)] \ .
		\end{equation}
	\end{cox}
	\noindent Note that there is no reason that we must take $a$ to be true first. The plausibility of $ab = ba$ could also be defined as
	\begin{equation}
		\mathcal{P}(ab) = g[\mathcal{P}(b),\mathcal{P}(a|b)] \ ,
	\end{equation}
	so that consistency requires
	\begin{equation}
		g[\mathcal{P}(a),\mathcal{P}(b|a)] = g[\mathcal{P}(b),\mathcal{P}(a|b)] \ .
	\end{equation}
	At this point, we have two unknown functions $f$ and $g$ that must be determined. In the next section, we will constrain the form of $g$ by applying it to special cases. Then in the following section, we will address the function $f$.
	
\section{The Product Rule}
	
	To further constrain the function $g$, Cox introduced an associativity consistency theorem. Suppose we wish to know the plausibility of the conjunction of three statements $abc$. Since $(ab)c = a(bc)$, Cox's second axiom implies
	\begin{equation}
		g[\mathcal{P}(ab),\mathcal{P}(c|ab)] = g[\mathcal{P}(a),\mathcal{P}(bc|a)] \ .
	\end{equation}
	The second axiom can be applied once more to the arguments to get
	\begin{equation}
		g\Big[g[\mathcal{P}(a),\mathcal{P}(b|a)],\mathcal{P}(c|ab)\Big] = g\Big[\mathcal{P}(a),g[\mathcal{P}(b|a),\mathcal{P}(c|ab)]\Big] \ .
	\end{equation}
	A functional equation of the form
	\begin{equation}
		g[g(x,y),z] = g[x,g(y,z)]
	\end{equation}
	has the general solution
	\begin{equation}
		g(x,y) = G^{-1}[G(x)G(y)] \ ,
	\end{equation}
	where $G$ is any invertible function \cite{cox:1946}.
	
	Applying $G$ to both sides of the general solution and replacing $x$ and $y$ with our plausibilities results in something remarkable,
	\begin{equation}
		G[\mathcal{P}(ab)] = G[\mathcal{P}(a)]G[\mathcal{P}(b|a)] \ ,
	\end{equation}
	We see that the plausibilities have become a product of the arbitrary $G$ functions. Note that the monotonic function $\mathcal{P}$ used to assign plausibilities is arbitrary. Since $G$ is also arbitrary and monotonic, we can simply `regraduate' our plausibilities by assigning a new arbitrary function $\mathsf{P}$ so as to simplify the associativity result,
	\begin{equation}
		\mathsf{P}(a) \eqdef G[\mathcal{P}(a)] \ ,
	\end{equation}
	so that the associativity result becomes
	\begin{equation}
		\label{eq:plaus_product_rule}
		\mathsf{P}(ab) = \mathsf{P}(a)\mathsf{P}(b|a) \ ,
	\end{equation}
	which is the familiar product rule of probability theory. This regraduation does not alter the ranking of the plausibilities but only changes the scale for the numbers we assign.

\section{Extreme Plausibilities}

	At this point, it is convenient to ask, what is the range of values assigned by $\mathsf{P}$? Since $\mathsf{P}$ is monotonic, the extremes of total certainty that a statement is true and total certainty that a statement is false should be assigned some unique, extreme numerical values $\mathsf{P}_T$ and $\mathsf{P}_F$, respectively. For any statement $a$, this means
	\begin{equation}
		\mathsf{P}(a|a) = \mathsf{P}_T \qquad \text{and} \qquad \mathsf{P}(a|\neg a) = \mathsf{P}(\neg a|a) = \mathsf{P}_F \ .
	\end{equation}
	We wish to find the most convenient choices $\mathsf{P}_T$ and $\mathsf{P}_F$.
	
	Consider the plausibility of the conjunction $ab$ when we know that $b$ is true. We assume that if $b$ is true, then the plausibility of $ab$ should be exactly the same as the plausibility of $a$,
	\begin{equation}
		\mathsf{P}(ab|b) = \mathsf{P}(a|b) \ .
	\end{equation}
	Using the product rule the plausibility of $ab$ given $b$ is
	\begin{equation}
		\mathsf{P}(ab|b) = \mathsf{P}(b|b)\mathsf{P}(a|bb) = \mathsf{P}_T\mathsf{P}(a|b) \ ,
	\end{equation}
	which implies
	\begin{equation}
		\mathsf{P}_T = 1 \ .
	\end{equation}

	Now consider the plausibility of the conjunction $a\neg b$ given that we know $b$ is true. Since we know that $b$ is true, the conjunction must be false, regardless of $a$,
	\begin{equation}
		\mathsf{P}(a\neg b|b) = \mathsf{P}_F \ .
	\end{equation}
	Using the product rule,
	\begin{equation}
		\mathsf{P}(a\neg b|b) = \mathsf{P}(a|b)\mathsf{P}(\neg b|b) = \mathsf{P}(a|b)\mathsf{P}_F \ ,
	\end{equation}
	so that
	\begin{equation}
		\mathsf{P}_F = \mathsf{P}(a|b)\mathsf{P}_F \ ,
	\end{equation}
	for all $a$. This condition only holds when $\mathsf{P}_F = 0$ or $\infty$. We are free to choose either range. For simplicity and for the sake of convention, we choose $\mathsf{P}_F = 0$ so that plausibilities lie in the usual range $[0,1]$ with 0 representing falsity and 1 representing truth. The choice of this particular range for the plausibilities is not unique, but simply a convenient choice.

\section{The Sum Rule}

	There is one final matter to resolve: the function $f$ from Cox's first axiom. Consider the plausibility of the conjunction $ab$,
	\begin{equation}
		\mathsf{P}(ab) = \mathsf{P}(a)\mathsf{P}(b|a) = \mathsf{P}(a)f[\mathsf{P}(\neg b|a)] \ .
	\end{equation}
	The conditional plausibility $\mathsf{P}(\neg b|a)$ can be replaced by noting that
	\begin{equation}
		\mathsf{P}(a\neg b) = \mathsf{P}(a)\mathsf{P}(\neg b|a) \ ,
	\end{equation}
	so that
	\begin{equation}
		\mathsf{P}(ab) = \mathsf{P}(a) f\left[ \frac{\mathsf{P}(a\neg b)}{\mathsf{P}(a)} \right] \ .
	\end{equation}
	Again, since $ab = ba$,
	\begin{equation}
		\label{eq:sum_rule_first}
		\mathsf{P}(a) f\left[ \frac{\mathsf{P}(a\neg b)}{\mathsf{P}(a)} \right] = \mathsf{P}(b) f\left[ \frac{\mathsf{P}(\neg ab)}{\mathsf{P}(b)} \right] \ .
	\end{equation}
	This expression must hold for all choices of $a$ and $b$. In particular, it most hold for the special case when $\neg b = ac$, for some $c$. It turns out that this special case constrains the form of $f$ greatly.
	
	We first simplify (\ref{eq:sum_rule_first}) by noting that
	\begin{equation}
		\mathsf{P}(a\neg b) = \mathsf{P}(aac) = \mathsf{P}(ac) = \mathsf{P}(\neg b) = f[\mathsf{P}(b)] \ .
	\end{equation}
	The plausibility $\mathsf{P}(\neg ab)$ in the right hand side of (\ref{eq:sum_rule_first}) is a bit more complicated. First recall from equation (\ref{eq:disjunction_negation}) that $\neg(a+b) = \neg a \neg b$. Then note that $\neg a \neg b = \neg a a c$ must be false. This further implies that $a + b$ must be true, which requires that either $a$ is true or $b$ is true. When $a$ is true, $\neg a = \neg a b$ must be false. When $b$ is true, we also have $\neg a b = \neg a$. Therefore, for this special case,
	\begin{equation}
		\mathsf{P}(\neg a b) = \mathsf{P}(\neg a) = f[\mathsf{P}(a)] \ .
	\end{equation}
	Making these substitutions, we get the following functional equation,
	\begin{equation}
		\mathsf{P}(a) f\left[ \frac{f[\mathsf{P}(b)]}{\mathsf{P}(a)} \right] = \mathsf{P}(b) f\left[ \frac{f[\mathsf{P}(a)]}{\mathsf{P}(b)} \right] \ .
	\end{equation}

	A functional equation of the form
	\begin{equation}
		x f\left[ \frac{f(y)}{x} \right] = y f\left[ \frac{f(x)}{y} \right]
	\end{equation}
	has the general solution
	\begin{equation}
		x^\alpha + [f(x)]^\alpha = 1 \ ,
	\end{equation}
	where $\alpha$ is a constant. Replacing $x$ with the plausibility of $a$ in this expression yields
	\begin{equation}
		[\mathsf{P}(a)]^\alpha + [\mathsf{P}(\neg a)]^\alpha = 1 \ .
	\end{equation}
	The utility of this result is apparent when we raise the product rule (\ref{eq:plaus_product_rule}) to the same power of $\alpha$,
	\begin{equation}
		[\mathsf{P}(ab)]^\alpha = [\mathsf{P}(a)]^\alpha [\mathsf{P}(b|a)]^\alpha \ .
	\end{equation}
	In one final regraduation, we can simplify the form of both of these expressions by defining a new function,
	\begin{equation}
		p(a) \eqdef [\mathsf{P}(a)]^\alpha \ ,
	\end{equation}
	so that we recover the standard sum and product rules of probability theory,
	\begin{gather}
		\label{eq:simple_sum_rule}
		p(a) + p(\neg a) = 1 \ , \\
		\label{eq:product_rule}
		p(ab) = p(a) + p(b|a) \ .
	\end{gather}
	We should note that the plausibility for a true statement and for a false statement are unchanged by this regraduation,
	\begin{equation}
		p_T = [\mathsf{P}_T]^\alpha = 1 \qquad \text{and} \qquad p_F = [\mathsf{P}_F]^\alpha = 0 \ .
	\end{equation}
	At this point, we see that the rules for assigning numbers to plausibilities in a consistent and convenient manner are the very same rules from probability theory. Therefore, we will simply refer to plausibilities as \emph{probabilities} from now on.

	The sum rule in (\ref{eq:simple_sum_rule}) can be used to derive a general sum rule stated in the following theorem:
	\begin{thm*}
		The probability of the disjunction $a+b$ is
		\begin{equation}
			p(a+b) = p(a) + p(b) - p(ab) \ .
		\end{equation}
	\end{thm*}
	\noindent To prove this, we recall equation (\ref{eq:disjunction_negation}) once more, $\neg(a+b) = \neg a\neg b$. Then using the sum rule (\ref{eq:simple_sum_rule}),
	\begin{align*}
		p(a+b) &= 1 - p(\neg a \neg b) \\
		&= 1 - p(\neg a)p(\neg b|\neg a) \\
		&= 1 - p(\neg a)[1 - p(b|\neg a)] \\
		&= 1 - p(\neg a) + p(\neg ab) \\
		&= p(a) + p(b)[1 - p(a|b)] \\
		&= p(a) + p(b) - p(ab) \ .
	\end{align*}

\section{Some Useful Consequences}
	
	Now that the cornerstones of probability theory are in place, we will comment on a few important consequences \cite{caticha:2008}. These consequences will be used extensively in the following chapters.

	\subsection{Independence}
	
		Two assertions are said to be \emph{independent} if knowledge of one does not affect the probability of the other. That is, if $a$ and $b$ are independent,
		\begin{equation}
			p(a|b) = p(a) \qquad \text{and} \qquad p(b|a) = p(b) \ .
		\end{equation}
		In this special case, the product rule (\ref{eq:product_rule}) simplifies to
		\begin{equation}
			p(ab) = p(a)p(b) \ ,
		\end{equation}
		and the sum rule (\ref{eq:simple_sum_rule}) becomes
		\begin{equation}
			p(a+b) = p(a) + p(b) - p(a)p(b) \ .
		\end{equation}

	\subsection{Mutual Exclusivity}
	
		Two statements are said to be \emph{mutually exclusive} if they cannot be simultaneously true. For two statements $a$ and $b$,
		\begin{equation}
			p(ab) = 0 \ ,
		\end{equation}
		even if $p(a) \ne 0$ and $p(b) \ne 0$. In this special case, the sum rule (\ref{eq:simple_sum_rule}) simplifies to
		\begin{equation}
			p(a+b) = p(a) + p(b) \ .
		\end{equation}
		The generalization to $n$ mutually exclusive assertions is straightforward,
		\begin{equation}
			p(a_1 + a_2 + \cdots + a_n) = \sum^{n}_{i=1} p(a_i) \ .
		\end{equation}

		A list of statements $a_1,a_2,\cdots,a_n$ is said to be \emph{exhaustive} if one or more of the statements must be true in any given situation. That is,
		\begin{equation}
			p(a_1 + a_2 + \cdots + a_n) = 1 \ .
		\end{equation}
		If the list of $n$ statements are both mutually exclusive \emph{and} exhaustive, this implies
		\begin{equation}
			\sum^{n}_{i=1} p(a_i) = 1 \ .
		\end{equation}
		This result holds regardless of whether the statements $a_1,a_2,\cdots,a_n$ are conditional on some statement $b$,
		\begin{equation}
			\label{eq:sum_of_conditional}
			\sum^{n}_{i=1} p(a_i|b) = 1 \ .
		\end{equation}

	\subsection{Marginalization}
	
		Consider a list of mutually independent assertions $a_1,a_2,\cdots,a_n$. Assume we only know the joint probabilities $p(a_ib)$, where $b$ is some other statement. How do we determine the probability of the assertion $b$ alone? The process, called \emph{marginalization}, is a straightforward consequence of (\ref{eq:sum_of_conditional}),
		\begin{equation}
			\sum^n_{i=1} p(a_ib) = p(b)\sum^{n}_{i=1} p(a_i|b) = p(b) \ .
		\end{equation}
		This procedure is tremendously useful when the assertions $a_i$ are unknown or not relevant.

\section{Expectation Values}
	
	One final concept in probability theory is needed: the notion of mean or expected values. If a variable $x$ can take the values $x_i$ with corresponding probability $p_i$, then the expectation value of some function of $x$ is defined as
	\begin{equation}
		\langle f(x) \rangle \eqdef \sum_i p_i f(x_i) \ .
	\end{equation}
	This expected value need not be an allowed value for $f(x)$, but is just a convenient estimate.
	
	The \emph{variance} of a function of the variable $x$ is defined as
	\begin{equation}
		\Delta f(x) \eqdef \left\langle (f(x) - \langle f(x)\rangle)^2 \right\rangle \  = \langle f^2(x) \rangle - \langle f(x) \rangle^2,
	\end{equation}
	and gives an estimate of the deviations from the mean value $\langle f(x)\rangle$. The \emph{standard deviation} is defined as the square-root of the variance, $(\Delta f(x))^{1/2}$ \cite{caticha:2008}.

\section{Continuous Probabilities}

	Up to this point we have discussed probabilities when variables only take a finite number of values. If a variable $X$ is continuous, however, there are \emph{infinitely} many possible values. In this case, the probability of one particular outcome is rather useless,
	\begin{equation}
		p(X{=}x) = 0 \ .
	\end{equation}
	A more useful quantity is the probability that the value of $X$ lies in the range $(x, x+dx)$,
	\begin{equation}
		p(x < X < x+dx) = \rho(x) \, dx \ ,
	\end{equation}
	where $\rho(x)$ is called the \emph{probability density}. The probability for $X$ to lie in the range $(a,b)$ is determined by integrating,
	\begin{equation}
		p(a < X < b) = \int^a_b\!dx\ \rho(x) \ .
	\end{equation}
	The probabilities are, of course, normalized so that
	\begin{equation}
		p(-\infty < X < \infty) = \int^\infty_{-\infty}\!dx\ \rho(x) = 1 \ .
	\end{equation}
	
	A uniform distribution is one that assigns equal probabilities to equal volumes. If a variable $x$ is defined in Cartesian coordinates, the choice is obvious, $\mu(x) = \text{constant}$. In a curved space, however, the volume elements are determined by the metric $g_{ab}$ of the space. In this case, the uniform probability should be $\mu(x) \propto g^{1/2} = \det(g_{ab})^{1/2}$.
	
	When dealing with continuous variables, one is almost always referring to probability densities. When there is no risk of confusion, it is common practice to refer to probability densities as simply probability distributions or even just probabilities. For simplicity, our arguments will be written in terms of continuous probability distributions whenever possible. The translation into discrete probabilities is straightforward,
	\begin{equation}
		p(x)\ dx \quad \rightarrow \quad p_i \ ,
	\end{equation}
	and
	\begin{equation}
		\int\!dx\ p(x) \quad \rightarrow \quad \sum_i p_i \ .
	\end{equation}

\section{Conclusions}

	In this chapter we have shown that it is possible to quantify a state of knowledge by assigning the plausibility of statements in a consistent and rational way. The resulting formalism takes the form of the very familiar rules of probability theory. We continue the theme of consistency and rationality in the next chapter where we examine how to update probabilities.

\mychapter{Updating Probabilities}
	\label{chapter:updating_probs}
	\epigraph{A wise man changes his mind, a fool never.}{A Spanish Proverb}

\noindent We showed in the previous chapter how one can assign the plausibility of statements in a consistent, rational way. The plausibilities assigned by a rational agent represents their state of knowledge. If the state of knowledge changes when information is acquired, the probabilities representing that state of knowledge must update accordingly. The goal of this chapter is to describe a means of updating probabilities in an objective way. Along the way, we demonstrate that statistical mechanics is simply an example of inference that arises when one attempts to make predictions with incomplete information.

\section{Bayes' Theorem}
	\label{sec:up_bayes}
	
	The first method used to update probabilities, Bayes' theorem, allows one to update their state of knowledge when data is acquired. Bayes' theorem is named for Reverend Thomas Bayes who originally developed the method over 300 years ago. The modern form of Bayes' theorem is actually due to the great mathematician Pierre-Simon Laplace. Laplace developed his version of the theorem independently and applied it to problems in numerous fields \cite{sivia:2006}.
	
	Consider some statements $\theta$. We wish to know how to update the probabilities representing our knowledge of $\theta$ when we learn some data $D$, where $D$ could be a set of any number of data values. If our knowledge before accounting for data is the \emph{prior} probability $q(\theta)$, what should our \emph{posterior} probability $p(\theta)$ be after the data is accounted for? We start by examining the joint prior $q(\theta,D)$. We can rewrite this joint prior with the product rule as
	\begin{equation}
		q(\theta,D) = q(\theta)q(D|\theta)
	\end{equation}
	or
	\begin{equation}
		q(\theta,D) = q(D)q(\theta|D) \ .
	\end{equation}
	Equating these two results and rearranging gives us Bayes' theorem,
	\begin{equation}
		\label{eq:bayes_theorem}
		q(\theta|D) = q(\theta)\frac{q(D|\theta)}{q(D)} \ .
	\end{equation}
	The probability $q(D|\theta)$ is known as the \emph{likelihood} as it represents the likelihood that the statements $\theta$ would generate the data $D$. The denominator $q(D)$ is a normalization factor called the \emph{evidence} that represents the probability of collecting the data $D$ regardless of the value of $\theta$.
	
	The power of this theorem is immediate when combined with the following rule, known as Bayes' rule. It states that the posterior distribution $p(\theta)$ which should be chosen in order to reflect the inclusion of information from data is simply the prior probability for $\theta$ conditional on $D$,
	\begin{equation}
		p(\theta) = q(\theta|D) \ .
	\end{equation}
	This rule is perhaps so intuitive and so obvious that it is typically overlooked, but aside from being reasonable, \emph{why} should we choose $q(\theta|D)$ as our posterior? This question will be addressed later in this chapter in section \ref{section:up_statmech}.

\section{Entropy}
	We now turn our attention to the concept of entropy. As we shall see, entropy plays a crucial role in updating probabilities.
	
	The concept of entropy has a long and complicated history \cite{flamm:1998}. It began as a purely thermodynamic quantity when, in 1865, Rudolf Clausis introduced \emph{thermodynamic entropy} as a useful concept in describing the behavior of heat engines.
	
	In the following years, thermodynamics was given a statistical interpretation by Boltzmann, founding \emph{statistical mechanics}. The key result of Boltzmann's work was that systems in thermodynamic equilibrium are governed by the Boltzmann distribution,
	\begin{equation}
		\label{eq:boltzmann_dist}
		p_i = \frac{e^{-\beta E_i}}{Z} \ ,
	\end{equation}
	where $Z$ is a normalization factor called the partition function, $E_i$ is the energy of an $N$-particle state, $\beta = 1/k_BT$, $k_B$ is the Boltzmann constant, and $T$ is the temperature of the system. For this probability distribution, it was found that the following definition known as the \emph{Gibbs entropy},
	\begin{equation}
		\label{eq:gibbs_entropy}
		S = -k_B \sum_i p_i \log p_i \ ,
	\end{equation}
	 coincided with the thermodynamic entropy \cite{jaynes:1965}. Although given a statistical interpretation, entropy was regarded as a purely thermodynamical concept. (It is still viewed in this way by many today.)

	Everything changed with Claude Shannon in 1948 \cite{shannon:1948}. Shannon was attempting to measure information loss in phone signals and wanted to develop a function that measured this `uncertainty.' He achieved his goal by asserting a few axioms that this measure would have to obey. We will not reproduce his arguments here, but his resulting measure of missing information is
	\begin{equation}
		\label{eq:shannon_entropy}
		S = -k \sum_i p_i \log p_i \ ,
	\end{equation}
	for some constant $k$, which has the same form as the Gibbs entropy in (\ref{eq:gibbs_entropy}). With encouragement from von Neumann, Shannon decided to call his measure `entropy.' Shannon states:
	\begin{quote}
		My greatest concern was what to call it. I thought of calling it `information,' but the word was overly used, so I decided to call it `uncertainty.' When I discussed it with John von Neumann, he had a better idea. Von Neumann told me, ``You should call it entropy, for two reasons. In the first place your uncertainty function has been used in statistical mechanics under that name, so it already has a name. In the second place, no one knows what entropy really is, so in a debate you will always have the advantage.' \cite{tribus:1971}
	\end{quote}
	As von Neumann pointed out, Shannon's \emph{rediscovery} of entropy in an entirely unrelated context only underscored the general confusion that existed about entropy and information in general.
	
	Shannon's derivation of entropy was remarkable as it was simply the result of arguments of consistency. Unfortunately, Shannon's work was limited to discrete probabilities only, and serious objections were later raised against Shannon's axioms and subsequent proofs \cite{uffink:1995}. Nevertheless, in the next section, we see how powerful Shannon's measure of ignorance can be. Later, in section \ref{sec:up_ME} we present a more general derivation of the concept of entropy.

\section{Statistical Mechanics and the Method of Maximum Entropy}
	\label{section:up_statmech}
	
	In 1952, L\'eon Brillouin made the claim that the similarities between the Gibbs entropy and Shannon's entropy were not a coincidence \cite{brillouin:1952}. He claimed that entropy is a general principle of inference.
	
	Brillouin's claims were realized in 1957 when Edwin Jaynes provided a ground-breaking derivation of statistical mechanics \cite{jaynes:1957}. In Jaynes' derivation, entropy is a tool from which the Boltzmann distribution of statistical mechanics can be derived and not a \emph{consequence of} statistical mechanics as thermodynamic entropy was viewed.
	
	Jaynes work was built on what he called \emph{the principle of maximum entropy} or MaxEnt. The argument is as follows: if there are a number of possible probability distributions that satisfy the relevant information, you should pick the one that reflects the maximum ignorance about everything else. Since Shannon's entropy (\ref{eq:shannon_entropy}) is a measure of ignorance, one determines this most ignorant distribution by finding the one that maximizes the entropy. In practice, this maximization is performed using the method of Lagrange multipliers.
	
	Consider a simple example where all we know is that a variable $x$ can take $n$ discrete values. We wish to determine the probabilities $p_i$ so that the entropy (\ref{eq:shannon_entropy}) is a maximum. The only constraint is that the probabilities must be normalized, $\sum_{i=1}^n p_i = 1$. We maximize the entropy by introducing a Lagrange multiplier $\alpha$ for the normalization constraint,
	\begin{equation}
		\delta \Big[ S[p] - \alpha\tsum_{i=1}^n p_i \Big] = 0 \ .
	\end{equation}
	Varying the $p_i$'s to as to maximize the entropy implies,
	\begin{equation}
		-\sum_{i=1}^n (\log p_i + 1 + \alpha)\delta p_i = 0 \ .
	\end{equation}
	Assuming that the variations $\delta p_i$ are independent implies that the distribution that maximizes the entropy is
	\begin{equation}
		p_i = e^{-1 - \alpha} \ .
	\end{equation}
	Since $\alpha$ is a constant, normalization implies
	\begin{equation}
		\tsum_{i=1}^n e^{-1-\alpha} = n e^{-1-\alpha} = 1 \ ,
	\end{equation}
	so that
	\begin{equation}
		p_i = \frac{1}{n} \ .
	\end{equation}
	If the only information we have about the probabilities is that they must be normalized, the distribution that reflects the most ignorance is a uniform distribution.
	
	Following Jaynes \cite{jaynes:1957}, we now consider the case where we have information about the expected values of some functions $f_r(x)$,
	\begin{equation}
		\label{eq:maxent_canon_contraints}
		\langle f_r(x) \rangle = \tsum_i p_i f_r(x_i) = F_r\ .
	\end{equation}
	We vary the $p_i$'s to maximize the entropy subject to normalization and the constraints on the expectation values,
	\begin{equation}
		\delta \Big[ S[p] - \alpha\tsum_i p_i - \tsum_r\lambda_r \left(\tsum_i p_i f_r(x_i)\right) \Big] = 0 \ ,
	\end{equation}
	which implies
	\begin{equation}
		p_i = \frac{1}{Z}\exp\left[- \tsum_r \lambda_r f_r(x_i)\right] \ ,
	\end{equation}
	where the partition function,
	\begin{equation}
		Z = \sum_i\exp\left[- \tsum_r \lambda_r f_r(x_i)\right]
	\end{equation}
	is a constant that comes from the normalization condition. The selected $p$ technically only ensures that the entropy is an extremum. It is straightforward to show, however, that this choice does yield a maximum \cite{caticha:2008}. The Lagrange multipliers are determined by comparing the selected distribution to the constraints in (\ref{eq:maxent_canon_contraints}).

	The application to statistical mechanics is immediate. If the only relevant information is that the expected value of the energy has some value,
	\begin{equation}
		\langle E \rangle = \tsum p_i E_i = \bar{E} \ ,
	\end{equation}
	the distribution that maximizes the entropy is the \emph{canonical distribution},
	\begin{equation}
		p_i = \frac{ e^{-\beta E_i}}{Z} \ .
	\end{equation}
	For an ideal gas, the canonical distribution is simply the Boltzmann distribution (\ref{eq:boltzmann_dist}). The conclusion is that statistical physics is nothing more than an example of inference where the expected value of the energy is the relevant information.

\section{The Extended Method of Maximum Entropy}
	\label{sec:up_ME}

	Shannon's goal was simply to quantify a loss of information; the resulting tool was entropy. Jaynes' goal was to determine maximally ignorant distributions subject to informational constraints; the resulting tool was again Shannon's entropy---specifically the method of maximum entropy. In his work, Jaynes demonstrated that information could come in the form of constraints.
	
	In section \ref{sec:up_bayes} we demonstrated that Bayes' theorem is a useful tool to update prior probabilities when information comes in the form of \emph{data}. However, in general, Bayes' theorem does not allow us to update priors when information comes in the form of \emph{constraints}. Many attempts have been made to determine a method of updating priors with constraints. In this section, we provide a derivation by Caticha \cite{caticha:2003,caticha:2007} that builds upon important contributions of the work of Shore and Johnson and the work of Skilling. (Shore and Johnson realized that one could axiomatize the updating method itself \cite{shore:1980}. Skilling's work contributes the idea that one needs a ranking of distributions in order of how much they update probabilities \cite{skilling:1988}.) In Caticha's derivation, he introduces a set of axioms that specify how the \emph{updating} process should behave. We will only present the axioms and jump to the conclusions as the proofs are lengthy.

	The idea here is simple: if you want to update from a prior distribution to a new posterior distribution when you gain information in the form of constraints, you should choose that posterior which is consistent with the acquired information but which changes your state of knowledge the least. This principle of minimal updating asserts that prior knowledge is important. Rather than throw away your prior when you learn something, you should try to stay as close to your prior as possible. As we will see, this minimal updating is achieved by maximizing the entropy.
	
	We seek some functional $S$ of the prior $q$ and possible posteriors $p$ that ranks the posteriors according to how much they update the prior. We appeal once more to Skilling's eliminative induction (section \ref{sec:quant_rational_plaus}) and assume that $S$ is sufficiently general. By applying a number of axioms and special cases, we hope to capture a single theory that describes the updating process. The axioms are as follows:
	
	\newtheorem{caticha}{Axiom}
	\begin{caticha}
		Local information has local effects.
	\end{caticha}
	\noindent Suppose a variable $x$ lies in the space $\mathcal{X}$. If the information we receive does not depend on a particular subdomain $\mathcal{D}$, then we do not want to update the probabilities in that subdomain, $p(x|\mathcal{D}) = q(x|\mathcal{D})$.
	
	\begin{caticha}
		The system of coordinates used contains no information.
	\end{caticha}
	\noindent We seek an updating process that is independent of the system of coordinates used. Coordinate systems are just labels that we use. If we decide to choose some other choice of labels, we do not expect our inferences to change.
	
	\begin{caticha}
		If a system is composed of independent subsystems, the updating process may treat them separately or jointly.
	\end{caticha}
	\noindent The implication is that if our prior is the probability of two independent variables $x_1$ and $x_2$,
	\begin{equation}
		q(x_1,x_2) = q_1(x_1)q(x_2) \ ,
	\end{equation}
	then it makes no difference if the joint prior $q(x_1,x_2)$ is updated or if the individual priors $q_i(x_i)$ are each updated independently. The result is that the posterior must also reflect the independence of $x_1$ and $x_2$,
	\begin{equation}
		p(x_1,x_2) = p_1(x_1)p(x_2) \ .
	\end{equation}

	Carrying out the implications of these axioms and applying the theory to special cases \cite{caticha:2003,caticha:2007}, results in only one functional $S[p,q]$ that ranks the possible posteriors $p$. The expression,
	\begin{equation}
		S[p,q] = -\int\!dx\ p(x) \log\frac{p(x)}{q(x)} \ ,
	\end{equation}
	is known as the \emph{relative entropy} (the entropy of $p$ relative to $q$). We recognize Shannon's entropy (\ref{eq:shannon_entropy}) as the special case when the variables are discrete and the prior is uniform. Since entropies are always relative to some distribution (uniform or otherwise), we will typically refer to the relative entropy as simply `the entropy.'
	
	Since the entropy $S[p,q]$ ranks the posteriors according to how they differ from the prior, we wish to choose that posterior which updates the prior the least. How do we know which $p$, ranked by $S$, updates the prior the least? It is obvious that when $p = q$, the entropy $S[q,q] = 0$. For any other $p$, we consider the concavity of the logarithm,
	\begin{equation}
		\log r \le r - 1 \ .
	\end{equation}
	Letting $r = q/p$ and multiplying both sides by $p$ gives
	\begin{equation}
		p(x) \log\frac{q(x)}{p(x)} \le q(x) - p(x) \ .
	\end{equation}
	Finally, if we invert the argument of the logarithm and take the integral of both sides, we see that
	\begin{equation}
		S[p,q] \le 0 \ ,
	\end{equation}
	with the equality holding only when $p = q$. The entropy ranks all posteriors by how much they update the prior, and the entropy of all posteriors is \emph{less} than the entropy of $q$ relative to itself. Therefore, to find the posterior that updates the prior the least, one must maximize the entropy. Jaynes' method of determining maximally ignorant distributions is then a special case where one is updating from a uniform prior and the probabilities are discrete. Accordingly, this more general process is called the \emph{extended method of maximum entropy} and is abbreviated ME.

\section{Unifying Bayes' Rule and the Method of Maximum Entropy}
	\label{section:up_unifying}
	
	At this point, we have two methods for updating probabilities. The first, Bayes' theorem is used when information comes in the form of \emph{data}. The second, the ME method, is used when information comes in the form of \emph{constraints}. For some time, it was not known whether these two methods were consistent or what their relationship was. In 2006, Caticha and Giffin demonstrated that, if the proper constraints were used, Bayes' theorem is shown to be a special case of ME \cite{caticha:2006}. We present their arguments here.

	As we saw earlier in section \ref{sec:up_bayes}, an inference problem involving data has two parts: a model, specified by parameters  $\theta$, generates possible data values $d$. Before performing an experiment, both of these are unknown. However, after the experiment has been performed, the data are the precisely known values $D$. We wish to know how the probability of the model parameters must update after our data has been collected.
	
	The appropriate distribution to update is the joint $p(\theta, d)$. We wish to find the particular $p(\theta,d)$ that maximizes the joint entropy,
	\begin{equation}
		S[p,q] = -\int\!d\theta\,dd\ p(\theta, d) \log \frac{p(\theta, d)}{q(\theta, d)} \ ,
	\end{equation}
	subject to constraints.\footnote{In this expression, $dd$ means the differential of the variable $d$.} In this problem, our constraints are normalization and the fact that the value of the data is known to be precisely the measured value $D$,
	\begin{equation}
		\label{eq:data_constraint}
		p(d) = \delta(d-D) \ ,
	\end{equation}
	so that the constraint on the joint posterior is
	\begin{equation}
		p(\theta,d) = \delta(d-D)p(\theta|d) \ .
	\end{equation}
	
	The condition on the data (\ref{eq:data_constraint}) introduces one Lagrange multiplier for each $d$, $\lambda(d)$. The posterior that maximizes the entropy is
	\begin{equation}
		\label{eq:bayes_posterior}
		p(\theta,d) = q(\theta,d)\frac{e^{\lambda(d)}}{Z} \ ,
	\end{equation}
	where $Z$ is a normalization constant. We can eliminate the Lagrange multiplier $\lambda(d)$ by integrating over $\theta$ and comparing with (\ref{eq:data_constraint}),
	\begin{equation}
		\int\!d\theta\ p(\theta,d) = q(d)\frac{e^{\lambda(d)}}{Z} = \delta(d-D) \ .
	\end{equation}
	Making this substitution in the posterior (\ref{eq:bayes_posterior}) yields
	\begin{equation}
		p(\theta,d) = \delta(d-D)q(\theta|d) \ .
	\end{equation}
	
	We are only concerned with the probability for the model parameters $\theta$, so we marginalize over the data $d$,
	\begin{equation}
		p(\theta) = \int\!dd\ \delta(d-D)q(\theta|d) = q(\theta|D) \ ,
	\end{equation}
	which can be recognized as Bayes' rule and can be written as Bayes' theorem,
	\begin{equation}
		p(\theta) = q(\theta|D) = q(\theta)\frac{q(D|\theta)}{q(D)} \ .
	\end{equation}

	The preceding result is satisfying for a number of reasons. First and foremost, it demonstrates that there is only one rule for updating probabilities---the extended method of maximum entropy. Bayes' theorem is simply a special case of the ME method when information comes as data. Second, it allows for generalizations of Bayes' theorem where we know \emph{both} data and constraints \cite{giffin:2007}.

	\subsection{Uncertainty and the Likelihood}

		The power of Bayes' theorem lies in the functional form of the likelihood $q(D|\theta)$. It is in this distribution that the relationship between the model parameters and the data is specified. In some cases this relationship is direct and straightforward. In more complicated problems, however, the specification of the likelihood can be considerably less obvious. In such cases, the likelihood is often constructed in an ad hoc fashion.

		In many cases, we can divide the contributions of the likelihood into two parts. The first part is the uncertainty introduced as an inherent feature of a probabilistic model. For example, in a problem involving coin flips or radioactive decay times, the outcomes are uncertain even if the model parameters (i.e. the coin bias and the decay constant, respectively) are known precisely. We call this type of uncertainty \emph{outcome uncertainty} as we cannot predict the outcomes precisely.
		
		The second type of uncertainty arises when we attempt to measure the outcome $x$ generated by a model with parameters $\theta$. The measurement process is not necessarily precise, and our possible collected data values $D$ may deviate from the true outcomes. We call this uncertainty \emph{data uncertainty} as the measured value of the data is uncertain given the outcome. A note of caution: many texts use the names `data' and `outcomes' interchangeably. Here they do not refer to the same things. We consider data $D$ as the result of a measurement of an outcome $x$, which is related but not identical.
		
		Bayes' theorem is rather indifferent to this distinction between uncertainties. In this way, the $x$ outcomes take on the character of intermediate or `nuisance' variables. However, the inclusion of this distinction is helpful to understand the various situations in which Bayes' theorem is applied.
		
		We can see how these two forms of uncertainty enter if we write the likelihood as a marginal over the joint distribution,
		\begin{equation}
			q(D|\theta) = \int\!dx\ q(x,D|\theta) \ .
		\end{equation}
		In problems where we can make such a distinction between uncertainties, the measurement devices measure outcomes of an experiment without regard for the model that produced it. In a freefall experiment, for example, a stopwatch simply measures the time of flight. It is indifferent to the value of the acceleration of gravity. So the data depends only on the outcomes, $q(D|\theta,x) = q(D|x)$. Our likelihood is then
		\begin{equation}
			q(D|\theta) = \int\!dx\ q(x|\theta)q(D|x) \ ,
		\end{equation}
		where we see the outcome uncertainty $q(x|\theta)$ and the data uncertainty $q(D|x)$. In a general problem where both uncertainties are present, this result shows how to systematically construct the likelihood. Just like Bayes' theorem itself, this result is sufficiently intuitive that it has been written in some form before \cite{gregory:2005}.
		
		As a final note, we will show how this likelihood simplifies to the two special cases in which Bayes' theorem is typically applied.

	\subsection{Precise Measurements}
		A precise measurement is one where there is no uncertainty in the data collected. That is, the data corresponds precisely to the outcomes,
		\begin{equation}
			q(D|x) = \delta(D - x) \ .
		\end{equation}
		One example of this type of problem is that of coin flips. Given the bias of the coin, the outcome of a flip is probabilistic, but after the flip there is no ambiguity in the measurement of which side the coin landed. In these problems, the likelihood simplifies to
		\begin{equation}
			q(D|\theta) = q(x{=}D|\theta) \ ,
		\end{equation}
		so that it is composed solely of outcome uncertainty.

	\subsection{Precise Outcomes}
		In many data analysis problems, the outcomes are given precisely by a deterministic model. The model is then a function of the parameters, $x=m(\theta)$. Knowledge of the parameters $\theta$ allows one to determine precisely the outcome,
		\begin{equation}
			q(x|\theta) = \delta(x - m(\theta)) \ .
		\end{equation}
		The uncertainty in such problems is solely due to the data uncertainty $q(D|x)$ that arises when we attempt to measure a precise outcome. If it were not for this data uncertainty, it would, in principle, only take a \emph{single} measurement to determine the model parameters. The likelihood in these problems is then
		\begin{equation}
			q(D|\theta) = q(D|x{=}m(\theta)) \ .
		\end{equation}
		If the data are independent and Gaussian distributed about the true outcome (as is often assumed\footnote{This can be determined by the ME method with the constraints $\langle d\rangle = x$ and $\langle (d - x)^2 \rangle = \sigma^2$. }),
		\begin{equation}
			q(D|x) = \prod_i p(D_i|x_i) \propto \prod_i \exp\left[ -\frac{(D_i-x_i)^2}{2\sigma^2} \right] \ ,
		\end{equation}
		then the likelihood is
		\begin{equation}
			q(D|\theta) \propto \exp\left[ \frac{1}{2\sigma^2}\sum_i (D_i - m_i(\theta))^2 \right] \ .
		\end{equation}
		If the prior is uniform or if there is a large amount of data collected, the most probable parameters of the posterior $p(\theta)$ are determined by \emph{maximizing the likelihood}. For Gaussian data uncertainty, this is achieved by \emph{minimizing the sum of the squares} in the previous expression, which is the origin of the method of least squares fitting.

\section{Systems of Constraints}
	
	Now that the ME method has been established as \emph{the} method for updating probabilities, we finish this chapter by pointing out different classes of systems of constraints. We will use figure \ref{fig:maxent_systems} to illustrate these different classes. In this figure we are depicting probability distributions as points in a `statistical manifold.' In general, this curved space has infinite dimensionality, but one can derive a unique metric on the space in order to talk about `distances' between two distributions \cite{murray:1993}.
	
	\subsection{Well-constrained Systems}
	
\begin{figure}
			\centering
			\subfloat[Well-constrained]{\label{fig:maxent_well}\includegraphics[width=0.45\textwidth]{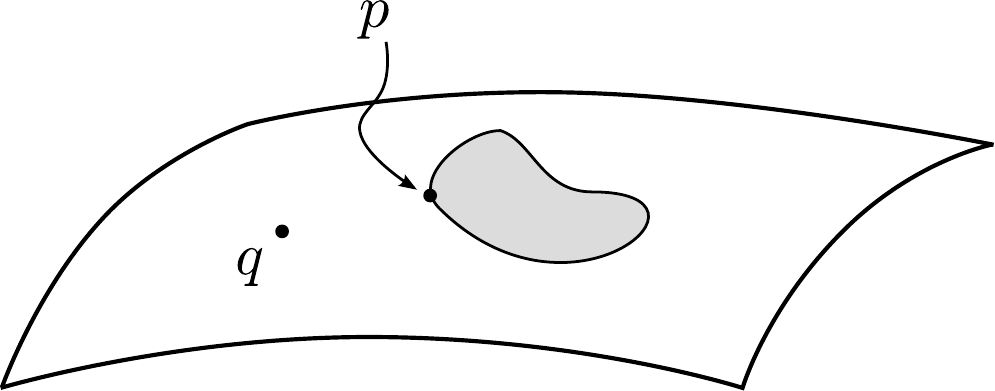}}                
			\quad
			\subfloat[Fully-constrained]{\label{fig:maxent_maximal}\includegraphics[width=0.45\textwidth]{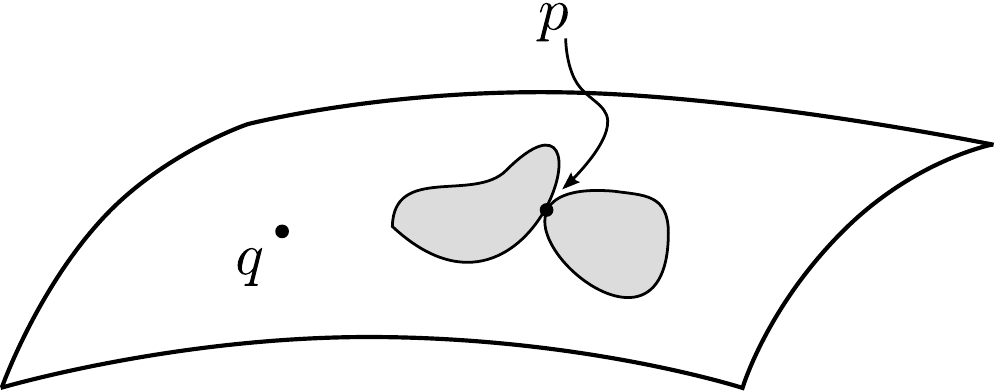}}
			\\
			\subfloat[Overconstrained]{\label{fig:maxent_over}\includegraphics[width=0.45\textwidth]{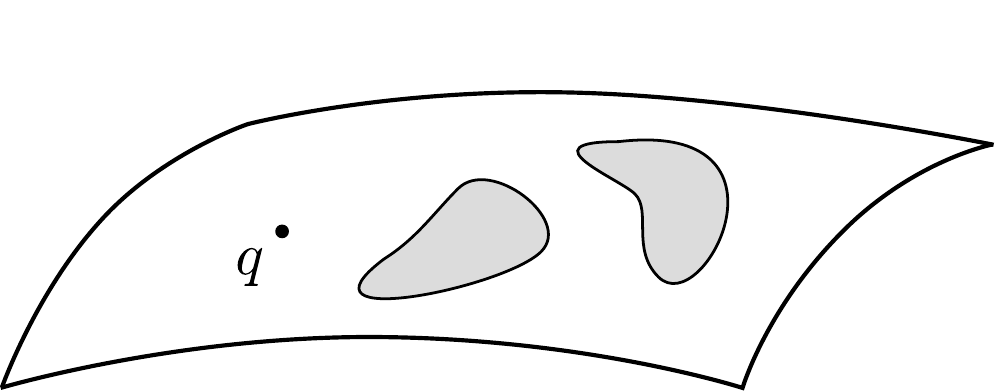}}                
			\quad
			\subfloat[Underconstrained]{\label{fig:maxent_under}\includegraphics[width=0.45\textwidth]{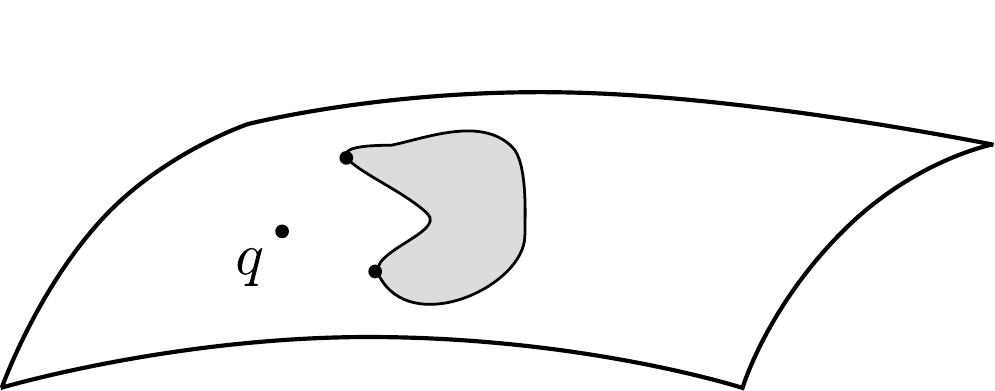}}
			\caption{Depictions of different classes of systems of constraints.}
			\label{fig:maxent_systems}
		\end{figure}

		The standard maximum entropy problem has a number of constraints and any number of possible posteriors that satisfy those constraints. The ME method then selects that posterior which maximizes the entropy. This is depicted in figure \ref{fig:maxent_well}. Many posterior distributions satisfy the constraints, but there is only one that maximizes the relative entropy $S[p,q]$ and, therefore, updates the prior the least.
		
	\subsection{Fully-constrained Systems}
	
		A system of constraints that is satisfied by only one possible posterior is said to be \emph{fully-constrained} (cf. figure \ref{fig:maxent_maximal}). The sole posterior $p$ automatically maximizes the entropy (as well as minimizing it) \emph{regardless} of the prior $q$. Consider updating from a prior $q(x)$ to a posterior $p(x)$ that satisfies the following constraints,
		\begin{equation}
			\langle x \rangle = \bar{x} \qquad \text{and} \qquad \Delta x = \left\langle (x - \bar{x})^2 \right\rangle = 0 \ .
		\end{equation}
		Provided $q(\bar{x}) \ne 0$, there is only one posterior distribution that can satisfy the constraints,
		\begin{equation}
			p(x) = \delta(x - \bar{x}) \ .
		\end{equation}
		In fully-constrained problems, there is often no need to \emph{explicitly} employ the ME method. Simply solving the constraint equations is sufficient.

	\subsection{Overconstrained Systems}
	
		An \emph{overconstrained} system is one in which \emph{no} posteriors satisfy the constraints. This is depicted in figure \ref{fig:maxent_over}. The implication in such a problem is that there is no way to rationally assign probabilities that are consistent with the constraining information---the information is \emph{incompatible}. Fully-constrained and overconstrained systems will be of great importance when we discuss quantum measurement in chapter \ref{chapter:measurement}.
	
	\subsection{Underconstrained Systems}
	
		The last class of constraints, \emph{underconstrained} systems, is shown in \ref{fig:maxent_under}. In these problems, the information is insufficient to determine a \emph{unique} posterior that maximizes the entropy for a given prior.
		
		For example, consider updating the uniform prior $\mu(x) \propto g^{1/2}$ with the constraint that the variance
		\begin{equation}
			\label{eq:under_var_constraint}
			\Delta x = \left\langle (x - \bar{x})^2 \right\rangle = \sigma^2 \ ,
		\end{equation}
		where we learn no information about the mean $\bar{x}$. Maximizing the relative entropy subject to normalization and the constraint on the variance,
		\begin{equation}
 			\delta \left[ S + \alpha\int\!dx\ p(x) + \beta\int\!dx (x-\bar{x})^2 p(x) \right] = 0 \ ,
		\end{equation}
		yields the posterior
		\begin{equation}
			p(x) = \frac{1}{Z} \exp[-\beta(x-\bar{x})^2] \ .
		\end{equation}
		The variance of the posterior is $\Delta x = 1/2\beta$. Comparing this to the constraint on the variance (\ref{eq:under_var_constraint}) gives
		\begin{equation}
			p(x) = \frac{1}{\sqrt{2\pi\sigma^2}} \exp\left[ -\frac{(x-\bar{x})^2}{2\sigma^2} \right] \ ,
		\end{equation}
		for \emph{any} mean $\bar{x}$. Note that the maximized entropy,
		\begin{equation}
			S[p,\mu] = \int\!dx\ p(x) \log\frac{p(x)}{g^{1/2}} = \frac{1}{2} + \log\sqrt{2\pi\sigma^2} \ ,
		\end{equation}
		is independent of the choice of $\bar{x}$.
		
		The implication of underconstrained systems is the same in any maximization problem. If there are multiple solutions of a maximization problem, they are all equally valid. So for the purposes of inference, all of the posteriors are allowed, provided all relevant information is included.

\section{Conclusions}

	There are two key conclusions to draw from this chapter. The first and most important is that the extended method of maximum entropy (ME) is \emph{the} means of updating one's state of knowledge when information comes in the form of constraints. In the special case when the constraining information is data, the ME method reduces to Bayes' famous theorem. The many uses of Bayes' theorem follow naturally.
	
	The second conclusion is that Jaynes was able to use the method of maximum entropy to derive statistical mechanics. Prior to Jaynes work, statistical mechanics was thought to be somehow operating at the level of reality. His derivation, however, showed it to be operating at the epistemological level. Could there be other physical theories that are expressed more naturally in terms of information? We explore this question in the next chapter where we attempt to explain quantum theory as another example of inference.

\mychapter{Entropic Quantum Dynamics}
	\label{chapter:eqd}
	\epigraph{It seemed to me that the foundation of the work of the mathematical physicist is to get the correct equations, that the interpretation of those equations was only of secondary importance.}{P. A. M. Dirac \cite{dirac:1997}}

\noindent With the successful application of informational principles to statistical mechanics by Jaynes, the natural question is, what other physical theories can be cast in terms of information? In 2009, A. Caticha provided a derivation of quantum mechanics with the method of maximum entropy taking center-stage \cite{caticha:2009}. The theory, \emph{entropic quantum dynamics} (EQD), is the subject of this chapter.

It should be stated upfront that the goal of entropic dynamics is not to supplant quantum mechanics with some entirely different theory or some more fundamental unified theory. We wish to do for quantum mechanics what Jaynes did for statistical mechanics. By interpreting statistical mechanics as an inference problem, the quantitative predictions of the theory remained unchanged, but the \emph{meaning} changed substantially. Likewise, we seek to reinterpret quantum mechanics as an inference problem. The numerous details worked out in the last century remain the same. It is only their meaning that changes.

Quantum theory, as it stands today, has many conceptual difficulties that make finding such an alternative interpretation extremely important and not simply an academic interest. We begin this chapter by discussing some of these difficulties. Then we discuss some crucial differences in viewing a theory as physical versus informational, before moving on to the derivation of EQD. Finally, we point out fundamental differences between entropic dynamics and ``hidden variable theories.''

\section{The Search for an Alternative Quantum Theory}

	Quantum theory is perhaps the most successful and accurate physical theory to date \cite{mermin:1990}. It allows us to properly describe the behavior of solids, atomic spectral lines, blackbody radiation, and countless other phenomena that eluded description by other theories. Since its inception, however, quantum mechanics has been plagued with fundamental conceptual issues. P. A. M. Dirac, one of the great fathers of quantum theory, was so troubled by these problems that he once stated, ``It is because of these difficulties, I believe that the foundations of quantum mechanics has not been properly laid down.''\cite{pang:2005}
	
	While the quantitative predictions of quantum mechanics are not questioned, the \emph{meaning} of quantum mechanics remains a hotly debated issue. Numerous alternative theories have cropped up in order to dispel the troubling problems of the orthodox interpretation of quantum theory, but these alternatives frequently introduce additional difficulties and complexities.
	
	We present a brief survey of some conceptual issues of quantum mechanics. This list is by no means exhaustive, but serves to illustrate the motivation for an alternative quantum theory.
	\begin{itemize}
		\item \textit{Indeterminism} -- Quantum theory is fundamentally probabilistic. This stands in stark contrast to the classical realm where theories are deterministic. While many theories, like statistical mechanics, appear to have a random or probabilistic nature, they are still assumed to be driven by an underlying deterministic theory. Much work has been done to cast quantum mechanics in terms of such a ``hidden variable theory,'' where the randomness of quantum mechanics is due to an underlying, uncontrollable, and perhaps unknown deterministic theory \cite{greiner:2001,peres:1993}. Such theories and their relation to entropic dynamics will be discussed further in section \ref{section:eqd_hiddenvars}.
		\item \textit{The interpretation of the wave function} -- The primary object in quantum mechanics is the wave function $\Psi$. This complex function is said to describe the state of a quantum system. Its connection to reality is postulated by the Born Rule \cite{merzbacher:1997} where, for example, the probability density for a particle's position is given by
		\begin{equation}
			\rho(x,t) = |\Psi(x,t)|^2 \ .
		\end{equation}
		It is important to note that the wave function describes an \emph{individual} particle or system. However, it only manifests itself in the \emph{ensemble} where the wave function's probabilistic nature is revealed \cite{bohm:1966}.
		\item \textit{Locality} -- Quantum mechanics has non-local consequences. It is experimentally verifiable that so-called `entangled' particles have correlations even when separated by an arbitrary distance \cite{greiner:2001}. This is at odds with our macroscopic experience, and the notion of instantaneous action across distances is not compatible with relativity, where simultaneity no longer exists \cite{peres:1993}.
		\item \textit{The measurement problem} -- Perhaps the most stubborn difficulty in quantum theory is that of measurement, where there is still no generally accepted solution \cite{kibble:1978}. In essence, the problem lies in the connection between the microscopic world where probabilities reign and the rigid, deterministic macroscopic world \cite{zurek:2002}. This problem, and the way in which entropic dynamics handles it, will be discussed in detail in chapter \ref{chapter:measurement}.
	\end{itemize}
	These subtle conceptual problems in quantum theory are often overlooked because of the wildly successful predictions it makes.

\section{Information and Reality}

	Entropic quantum dynamics, like Jaynes' approach to statistical mechanics, operates at the epistemological level. That is, it is formulated in terms of information, not reality. This distinction is subtle, but crucial.
	
	The goal of physics is to order and interpret reality \cite{stapp:1972,johnson:2010}. This goal is only successful if nature adheres to some logical ordering. We refer to this ordering as \emph{the laws of nature}. In turn, we collect observations about reality and attempt to find the relationship between these observations. These relationships are \emph{the laws of physics} and take a mathematical form. The laws of physics describe the \emph{information} we have about reality, not necessarily reality itself.
	
	The standard viewpoint is that the laws of physics and the laws of nature are one and the same or that, at best, the laws of physics approximate the laws of nature. In fact, this view is so ingrained that many cannot conceive of a distinction. This view may, in fact, be true---reality may really consist of particles, strings, wave functions, or any other macroscopic description we may use. However, it may be that the connection is far more complicated. H. P. Stapp writes:
	\begin{quote}
		The proper goal of science is to augment and order our experience. A scientific theory should be judged on how well it serves to extend the range of our experience and reduce it to order. It need not provide a mental or mathematical image of the world itself, for the structural form of the world itself may be such that it cannot be placed in simple correspondence with the types of structures that our mental processes can form.\cite{stapp:1972}
	\end{quote}
	In this case, the rules for processing information become highly relevant.
	
	If our physical theory describes information, the powerful rules for processing information presented in chapters \ref{chapter:quantifying_uncertainty} and \ref{chapter:updating_probs} can be used. As seen with statistical mechanics in section \ref{section:up_statmech} and with quantum mechanics later in this chapter, these powerful rules can greatly simplify the problem at hand.
	
	In the next section, we show how these powerful rules of inference can be applied to a quantum system. However, a particular alternative approach to quantum mechanics, \emph{stochastic mechanics}, by Edward Nelson should be singled out first. In 1966, Nelson formulated a derivation of quantum mechanics that attempted to explain quantum phenomena in classical Newtonian terms \cite{nelson:1966,nelson:1985}. Stochastic mechanics is a hidden variable theory that assumes that there is some ``background field'' that causes fluctuations and leads to a form of non-dissipative Brownian motion. The theory introduced many additional conceptual problems and complexity, which ultimately lead Nelson to abandon the theory \cite{nelson:1986}. One such problem is discussed in section \ref{sec:measurement_sequential}.
	
	On the surface, stochastic mechanics and entropic quantum dynamics are very similar. They share some of the same assumptions and, in some ways, are formally very similar. The major distinction between the two theories, however, is the very subject of this section---Nelson's arguments were ontological in nature. Stochastic mechanics claims to describe reality itself, whereas entropic dynamics is attempting to describe the limited information we possess about microscopic systems. As a result, EQD is much less complex and side-steps many of the conceptual difficulties that arise in stochastic mechanics.

\section{The Statistical Model in EQD}
	\label{sec:eqd_model}
	
	In this section we present the derivation of entropic quantum dynamics by Caticha in \cite{caticha:2009,caticha:2010}.
	
\begin{figure}
		\centering
		\includegraphics[scale=0.75]{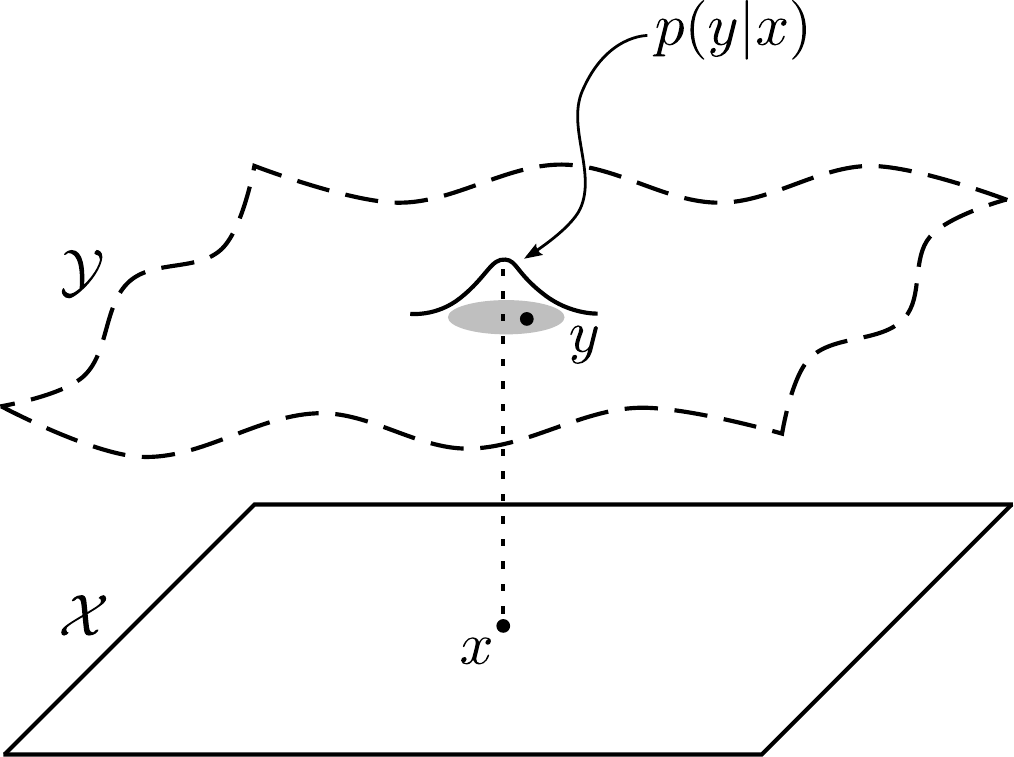}
		\caption[Relationship between the position of a particle $x$ and the extra variables $y$]{\label{manifolds} The position of a particle is a point $x$ in the flat, three-dimensional configuration space $\mathcal{X}$. For each $x$ there exists a probability distribution $p(y|x)$ for the extra variables associated with the particle.}
	\end{figure}

	We consider a particle at a position $x$ in a flat, three-dimensional configuration space $\mathcal{X}$. Associated with the particle are a number of hidden variables living in a space $\mathcal{Y}$, denoted $y$. These extra `$y$-variables' have some uncertainty that depends on the position of the particle. Accordingly, this uncertainty is described by a probability distribution, $p(y|x)$ (cf. Figure \ref{manifolds}). Remarkably, we need not specify the form of $p(y|x)$. 
	
	The metric in $\mathcal{X}$ is flat but scaled by a small variance,
	\begin{equation}
		\label{metric}
		\gamma_{ab} = \frac{\delta_{ab}}{\sigma^2} \ .
	\end{equation}
	The small scale factor is included to allow generalization to multiple particles. For $N$ particles, the configuration space $\mathcal{X}_N$ would be $3N$-dimensional, flat but anisotropic. The metric for 2 particles, for example, would be
	\begin{equation}
		\gamma_{AB} = \left[ \begin{array}{cc}
		        \delta_{a_1b_1}/\sigma_1^2 & 0 \\
		        0 & \delta_{a_2b_2}/\sigma_2^2
		       \end{array} \right] \ ,
	\end{equation}
	where each $\delta_{a_ib_i}$ is a $3\times3$ matrix. The choice of different scale factors leads to particles with different masses.

\section{Introducing Dynamics}

	We must introduce another basic assumption: small changes from one state to another happen and, furthermore, that large changes are the accumulation of many small changes.
		
	Consider a particle at an initial position $x$ moving to some unknown position $x'$. The only information we have about the particle's final position is that the difference in positions must be small. The best we can hope for is a probability of a final position. To find this probability, we will apply the method of maximum entropy subject to normalization and the constraint that steps from one position to another must be small.
	
	When a particle moves, not only are we ignorant of its final position $x'$, but we are also ignorant of the corresponding extra variables $y'$ in the new state. Therefore, the relevant space for our problem is $\mathcal{X} \times \mathcal{Y}$, where we wish to find the distribution $P(x', y'|x)$. The relative entropy that must be maximized is
	\begin{equation}
		\label{eq:eqd_relative_entropy}
		\mathcal{S}[P,Q] = -\int \! dx' dy' \ P(x',y'|x) \log \frac{P(x',y'|x)}{Q(x',y'|x)} \ .
	\end{equation}
	We assume that the prior $Q(x',y'|x)$ represents a state of complete ignorance. That is, $x'$ and $y'$ are independent of each other and uniform with respect to their own volume elements. Therefore, the joint prior is
	\begin{equation}
		Q(x',y'|x) = Q(x'|x)Q(y'|x) = \gamma^{1/2}q(y) \ ,
	\end{equation}
	where $\gamma = \det( \gamma_{ab})$ and $q(y)$ is a uniform measure of the space $\mathcal{Y}$.
	
	Apart from normalization, there are two additional constraints on the posterior. First, we require that $x'$ and $y'$ be related by $p(y'|x')$. Second, we require that the new variables $y'$ depend only on the new position $x'$. That is,
	\begin{equation}
		P(x',y'|x) = P(x'|x)p(y'|x') \ ,
	\end{equation}
	so that the $y'$ variables are independent of the prior position.
	Finally, we impose the constraint that $x'$ be close to $x$. We require that
	\begin{equation}
		\label{eq:small_step_constraint}
		\langle \Delta \ell^2(x',x)\rangle = \langle \gamma_{ab} \Delta x^a \Delta x^b \rangle = \Delta \bar{\ell}^2 (x) \ ,
	\end{equation}
	where $\Delta x^a = x'^a - x^a$ and $\Delta \bar{\ell}^2 (x)$ is some small and, for now, unspecified value.

	Now that the stage has been set, we apply the method of maximum entropy. The probability distribution that maximizes $\mathcal{S}[P,Q]$ subject to the constraints is
	\begin{equation}
		\label{eq:raw_prob}
		P(x'|x) = \frac{1}{\zeta(x, \alpha)} \exp \left[ S(x') - \frac{1}{2}\alpha(x) \Delta\ell^2(x',x) \right] \ ,
	\end{equation}
	where $\zeta(x, \alpha)$ is a normalization factor, and $\alpha(x)$ is a Lagrange multiplier that controls the step size. We see the a large $\alpha$ leads to small steps, while a small $\alpha$ leads to large ones. $S(x)$ is the entropy of the $y$ variables relative to an underlying measure $q(y)$ of the space $\mathcal{Y}$,
	\begin{equation}
		\label{entropy}
		S(x) = -\int\! dy \ p(y|x) \log \frac{p(y|x)}{q(y)} \ .
	\end{equation}
	We cannot evaluate the $y$ variable entropy directly without knowing the distribution $p(y|x)$ or the nature of the $y$ variables themselves. As we will show, however, we can simply work at the level of this \emph{entropy field}.
	
	For large $\alpha$, we can write $x'^a = x^a + \Delta x^a$. We can then expand the exponent of the transition probability (\ref{eq:raw_prob}) about its maximum to get
	\begin{equation}
		\label{eq:approx_trans_prob_w_alpha}
		P(x'|x) \approx \frac{1}{Z(x)} \exp \left[ -\frac{\alpha(x)}{2\sigma^2} \delta_{ab} (\Delta x^a - \Delta \bar{x}^a)(\Delta x^b - \Delta \bar{x}^b) \right] \ ,
	\end{equation}
	where $Z(x)$ is a new normalization factor. The displacement $\Delta x^a$ can be expressed as the expected displacement plus a fluctuation,
	\begin{equation}
		\Delta x^a = \Delta \bar{x}^a + \Delta w^a \ ,
	\end{equation}
	where
	\begin{equation}
		\langle \Delta x^a \rangle = \Delta \bar{x}^a = \frac{\sigma^2}{\alpha(x)} \partial^a S(x) \ ,
	\end{equation}
	\begin{equation}
		\label{eq:fluctuations_w_alpha}
		\langle \Delta w^a \rangle = 0 \quad \text{and} \quad \langle \Delta w^a \Delta w^b \rangle = \frac{\sigma^2}{\alpha(x)} \delta^{ab} \ .
	\end{equation}
	The particle drifts up the entropy gradient with $\alpha(x)$ controlling the step size. A larger $\alpha$ means smaller displacements. It should be noted that as $\alpha \rightarrow \infty$, the fluctuations dominate. As in Brownian motion, the trajectory is continuous, but non-differentiable.
	
\section{Time}

	Time is introduced as a means of keeping track of the accumulation of changes. For short steps, motion is described by the transition probability $P(x'|x)$ in (\ref{eq:approx_trans_prob_w_alpha}). Larger changes, on the other hand, are the accumulation of many short steps. Given an initial position $x$, the first step in the series is given by $P(x'|x)$. After the first step, however, we are uncertain of both $x'$ and $x$. We must deal with the joint probability $P(x', x) = P(x'|x)P(x)$. Integrating over $x$ gives us,
	\begin{equation}
		\label{eq:rho_def}
		P(x') = \int \! dx \ P(x'|x)P(x) \ ,
	\end{equation}
	the probability of the particle being at a particular point $x'$. If we interpret $P(x)$ to be the probability at a given time $t$ and $P(x')$ as the probability at a time $t' = t + \Delta t$, then we can write $P(x) = \rho(x,t)$ and $P(x') = \rho(x', t')$, giving us a notion of instants.
	
	Now that we have this notion of successive instants, we require the definition of the interval of time $\Delta t$ between them. Specifying this interval amounts to tuning the size of steps. The time that governs non-relativistic quantum mechanics, Newtonian time, flows equably at every point in space. That is, the interval of time $\Delta t$ must be uniform in space. This is achieved if $\alpha(x,t)$ is chosen as
	\begin{equation}
		\alpha(x,t) = \frac{\tau}{\Delta t} = \text{constant} \ ,
	\end{equation}
	where $\tau$ is a constant that ensures $\Delta t$ has units of time. This choice of a constant $\alpha$ reflects the translational symmetry present in the configuration space $\mathcal{X}$.
	
	Substituting this definition for $\alpha(x,t)$ simplifies the description of motion. In particular, the transition probability becomes
	\begin{equation}
		\label{eq:approx_trans_prob}
		P(x'|x) \approx \frac{1}{Z(x)} \exp \left[ -\frac{\tau}{2\sigma^2\Delta t} \delta_{ab} (\Delta x^a - \Delta \bar{x}^a)(\Delta x^b - \Delta \bar{x}^b) \right] \ ,
	\end{equation}
	with displacement
	\begin{equation}
		\label{eq:nt_displacement}
		\Delta x^a = b^a(x)\Delta t + \Delta w^a \ .
	\end{equation}
	The drift velocity $b^a(x)$ and fluctuation $\Delta w^a$ are given by
	\begin{equation}
		\label{eq:future_drift}
		\langle \Delta x^a \rangle = b^a(x)\Delta t \quad \text{where} \quad b^a(x) = \frac{\sigma^2}{\tau} \partial^a S(x) \ ,
	\end{equation}
	\begin{equation}
		\label{eq:nt_fluctuations}
		\langle \Delta w^a \rangle = 0 \quad \text{and} \quad \langle \Delta w^a \Delta w^b \rangle = \frac{\sigma^2}{\tau}\Delta t \delta^{ab} \ .
	\end{equation}
	The velocity in (\ref{eq:future_drift}) is identified as the mean velocity to the future.
	
	As a consequence of Bayes' theorem (\ref{eq:bayes_theorem}), this velocity to the future is not the same as the velocity from the past \cite{caticha:2009,caticha:2010}. While asymmetry in time is a natural consequence of inference, the Schr\"odinger equation, which is derived in the next section, turns out to be time-reversal invariant.

\section{The Schr\"odinger Equation} 
	\label{sec:eqd_schrod}

	The accumulation of the small steps in \ref{eq:rho_def} is described by a Fokker-Planck (FP) equation \cite{chandrasekhar:1943},
	\begin{equation}
		\partial_t \rho = - \partial_a(b^a \rho) + \frac{\sigma^2}{2\tau} \nabla^2 \rho \ .
	\end{equation}
	The first term arises from the expected displacement (\ref{eq:future_drift}), and the diffusion constant in the second term originates from the fluctuations in (\ref{eq:nt_fluctuations}). The Fokker-Plank equation can be rewritten as a continuity equation,
	\begin{equation}
		\label{eq:fp}
		\partial_t \rho = -\partial_a(v^a\rho) \ ,
	\end{equation}
	provided the \emph{current velocity} is
	\begin{equation}
		v^a = b^a - \frac{\sigma^2}{2\tau} \partial^a \log \rho \ .
	\end{equation}
	We can introduce an \emph{osmotic velocity} as
	\begin{equation}
		\label{eq:osmotic}
		u^a \overset{\underset{\mathrm{def}}{}}{=} - \frac{\sigma^2}{2\tau} \partial^a \log \rho \ ,
	\end{equation}
	such that $v^a = b^a + u^a$. The mean drift $b^a$ drives the probability flow up the entropy gradient while the osmotic velocity drags it down the concentration gradient, hence the name osmotic velocity.
	
	Since the current velocity $v^a$ is the sum of gradients, it can be written as a gradient as well,
	\begin{equation}
		\label{eq:current_v_grad_phi}
		v^a = \frac{\sigma^2}{\tau} \partial^a \phi \ ,
	\end{equation}
	where
	\begin{equation}
		\phi(x,t) = S(x) - \log \rho^{1/2}(x,t) \ .
	\end{equation}

	The dynamics described thus far is diffusion, not quantum mechanics. We must add one final ingredient. In order to construct a wave function $\Psi = \rho^{1/2}e^{i\phi}$, we need to promote the phase $\phi$ an independent degree of freedom by allowing the entropy field $S(x)$ to change in response to the dynamics.
	
	In order to specify the way in which $S(x,t)$ changes, we impose that the diffusion be conservative by requiring conservation of an energy functional $E[\rho, S]$. The notion of a \emph{conservative} diffusion was a prodigious idea introduced by Nelson in the development of stochastic mechanics \cite{nelson:1966,nelson:1985}. Our choice of energy constitutes extremely relevant information. At this point, we choose an energy that is quite reasonable, but further justification of this choice is a subject for future research.
	
	We choose a time-reversal invariant energy that is a functional of our velocities. As we mentioned earlier, the drift velocity $b^a$ from the past is not the same as to the future, so we will exclude it from our energy functional. On the other hand, under time reversal the current velocity $v^a \rightarrow -v^a$, and the osmotic velocity $u^a \rightarrow u^a$. In the low velocity limit, we only need to include terms of order $v^2$ and $u^2$,
	\begin{equation}
		\label{eq:energy1}
		E[\rho,S] = \int \! dx \ \rho(x,t) \left( \frac{1}{2}mv^2 + \frac{1}{2}\mu u^2 + V(x,t) \right) \ ,
	\end{equation}
	where $m$ and $\mu$ are constants that ensure the $E$ has units of energy. The constants are called the \emph{current mass} and \emph{osmotic mass}, respectively. The field $V(x,t)$ represents an external potential.
	
	It is convenient to define a new constant $\eta$ such that
	\begin{equation}
		\frac{\sigma^2}{\tau} = \frac{\eta}{m} \ .
	\end{equation}
	Substituting this new constant and the expressions for the velocities gives
	\begin{equation}
		\label{eq:energy2}
		E = \int \! dx \ \rho \left( \frac{\eta^2}{2m} (\partial_a\phi)^2 + \frac{\mu\eta^2}{8m^2} (\partial_a \log \rho)^2 + V \right) \ .
	\end{equation}
	We impose that the energy increase at the rate
	\begin{equation}
		\label{eq:energy_cons}
		\dot{E} = \int \! dx \ \rho \partial_t V \ .
	\end{equation}
	When the potential is time-independent (i.e.\ $\partial_t V = 0$), the energy conservation condition simplifies to
	\begin{equation}
		\label{eq:energy_cons_t_ind_V}
		\dot{E} = 0 \ .
	\end{equation}
	We can also write the FP equation (\ref{eq:fp}) in terms of the new constants,
	\begin{equation}
		\label{eq:continuity2}
		\partial_t \rho = -\frac{\eta}{m}(\partial^a\rho\partial_a\phi + \rho\nabla^2\phi) \ .
	\end{equation}

	If we take the time derivative of (\ref{eq:energy2}) and apply manipulations involving integration by parts and the FP equation (\ref{eq:continuity2}), we get
	\begin{equation}
		\dot{E} - \int \! dx \ \rho \partial_t V = \int\!dx \ \partial_t \rho\left[ \eta \partial_t \phi + \frac{\eta^2}{2m}(\partial_a\phi)^2 + V - \frac{\mu\eta^2}{2m^2} \frac{\nabla^2 \rho^{1/2}}{\rho^{1/2}} \right] \ .
	\end{equation}
	Imposing that the conservation condition hold for arbitrary choices in $\rho$ and $\phi$ implies
	\begin{equation}
		\label{eq:pde_phi}
		\eta \partial_t \phi + \frac{\eta^2}{2m}(\partial_a\phi)^2 + V - \frac{\mu\eta^2}{2m^2} \frac{\nabla^2 \rho^{1/2}}{\rho^{1/2}} = 0 \ ,
	\end{equation}
	which we call the \emph{phase equation}.

	Finally, we can combine the two coupled differential equations (\ref{eq:continuity2}) and (\ref{eq:pde_phi}) into a complex function
	\begin{equation}
		\label{eq:wavefunction}
		\Psi = \rho^{1/2} e^{i\phi} \ .
	\end{equation}
	This leads to the dynamical equation,
	\begin{equation}
		i\eta \partial_t \Psi = - \frac{\eta^2}{2m} \nabla^2\Psi + V\Psi + \frac{\eta^2}{2m}\left( 1-\frac{\mu}{m} \right) \frac{\nabla^2(\Psi\Psi^*)^{1/2}}{(\Psi\Psi^*)^{1/2}}\Psi \ .
	\end{equation}
	Identifying $\eta$ with $\hbar$ and the current mass with the osmotic mass, $m = \mu$, reproduces the Schr\"odinger equation,
	\begin{equation}
		\label{eq:schrod}
		i\hbar \frac{\partial\Psi}{\partial t} = -\frac{\hbar^2}{2m}\nabla^2\Psi + V\Psi \ .
	\end{equation}
	The choice of $m = \mu$ turns out to be a matter of tuning our units and our choice of the wave function (\ref{eq:wavefunction}) appropriately. One may always simplify the dynamics in this way \cite{caticha:2009,caticha:2010}.

\section{External Electromagnetic Fields}
	\label{sec:eqd_external_em}
	
	An external electromagnetic field can be accounted for in entropic dynamics by introducing an additional constraint. While the original constraint (\ref{eq:small_step_constraint}) moderates displacements in all directions, we introduce an external field that constrains displacements in particular directions,
	\begin{equation}
		\langle \Delta x^a n_a(x)\rangle = C(x) \ ,
	\end{equation}
	where $n_a(x)$ is the unit covector and $C(x)$ is the intensity of the external field. We introduce a new field that represents the magnitude of the external field in terms of the effect it induces,
	\begin{equation}
		A_a(x) \propto \frac{n_a(x)}{C(x)} \ ,
	\end{equation}
	so that the constraint becomes
	\begin{equation}
		\label{eq:field_constraint}
		\langle \Delta x^a A_a(x) \rangle = C \ ,
	\end{equation}
	where $C$ is a constant that reflects the strength of the coupling.
	
	With this additional constraint, the transition probability that maximizes the relative entropy (\ref{eq:eqd_relative_entropy}) is
	\begin{equation}
		P(x'|x) = \frac{1}{\zeta(x,\alpha,\beta)} \exp\left[ S(x') - \frac{1}{2}\Delta \ell^2(x',x) - \beta \Delta x^a A_a(x) \right] \ ,
	\end{equation}
	where $\beta$ is a Lagrange multiplier that comes from the additional field constraint and $\zeta$ is a normalization constant. We can once again expand the transition probability about the maximum displacement $\Delta\bar{x}^a$. Expressing a displacement as the expected plus a fluctuation gives
	\begin{equation}
		\Delta x^a = \Delta \bar{x}^a + \Delta w^a \ ,
	\end{equation}
	where
	\begin{equation}
		\langle \Delta x^a \rangle = \Delta \bar{x}^a = b^a\Delta t \qquad \text{with} \qquad b^a = \frac{\hbar}{m} [\partial^a S - \beta A^a] \ ,
	\end{equation}
	\begin{equation}
		\langle \Delta w^a \rangle = 0 \quad \text{and} \quad \langle \Delta w^a \Delta w^b \rangle = \frac{\hbar}{m} \delta^{ab} \ .
	\end{equation}
	We see that while the fluctuations remain unchanged, the drift velocity picks up an additional term coming from the external field.

	The small changes once again accumulate according to the Fokker-Planck equation (\ref{eq:fp}) but with a new current velocity,
	\begin{equation}
		v^a = \frac{\hbar}{m} [\partial^a S - \beta A^a] \ .
	\end{equation}
	The invariance of the fluctuations lead to an unchanged osmotic velocity (\ref{eq:osmotic}), and the phase $\phi$ is defined in the same way as (\ref{eq:current_v_grad_phi}).

	One again introducing energy conservation leads to a new phase equation,
	\begin{equation}
		\hbar \partial_t \phi + \frac{\hbar^2}{2m}(\partial_a\phi - \beta A_a)^2 + V - \frac{\hbar^2}{2m} \frac{\nabla^2 \rho^{1/2}}{\rho^{1/2}} = 0 \ .
	\end{equation}
	To determine the role of the Lagrange multiplier $\beta$, it is helpful to let $S_\text{HJ} = \hbar\phi$. In the classical limit as $\hbar \rightarrow 0$, the phase equation with this substitution reduces to the Hamilton-Jacobi equation for a particle in an external electromagnetic field provided
	\begin{equation}
		\beta = \frac{e}{\hbar c} \ ,
	\end{equation}
	where $e$ is the electric charge and $c$ is the speed of light in vacuum. Therefore, in EQD the electric charge originates as a Lagrange multiplier that the controls the response of a particle to the external potential $A_a$.
	
	Combining the phase equation with the Fokker-Plank equation so that $\Psi = e^{i\phi}$ results in the Schr\"odinger equation in the presence of an external electromagnetic field,
	\begin{equation}
		i\hbar \frac{\partial\Psi}{\partial t} = \frac{\hbar^2}{2m}\left( i\partial_a - \frac{e}{\hbar c} A_a \right)^2\Psi + V\Psi \ .
	\end{equation}
	The subject of gauge invariance in this derivation will be discussed in the next chapter on symmetry.

\section{A Remark on ``Hidden Variables''}
	\label{section:eqd_hiddenvars}

	Before concluding this chapter we wish to make a brief remark on the subject of ``hidden variable'' theories. Hidden variable theories were originally devised as a means of ascribing the probabilistic nature of quantum mechanics to some underlying, uncontrollable deterministic theory. The idea is that in the preparation of a system, it is not possible to control the preparation of some ``hidden'' degrees of freedom that influence the motion of the system. The `randomness' of quantum mechanics is only an illusion that appears because some element of reality is hidden.
	
	Faced with the inability to reconcile his theory of general relativity with quantum mechanics, Einstein became one of the biggest proponents of hidden variable theories, as evidenced by the now famous EPR paper \cite{einstein:1935}. However, in 1964 another famous paper by John Bell showed that \emph{local} hidden variable theories cannot explain all quantum phenomena \cite{bell:1964,bell:1966}. Quantum theories could be deterministic \emph{or} local, not both.
	
	Bell's result was a big blow to champions of the hidden variable mindset, but it did not rule out non-local hidden variable theories. Nelson's stochastic mechanics \cite{nelson:1966} and the de Broglie-Bohm pilot-wave theory \cite{bohm:1952} are two such examples of non-local hidden variable theories. These theories are far from the original notion of a hidden variable theory, but the goal is more or less the same: to explain the `strangeness' of quantum theory in terms of some simpler underlying theory. They have yet to succeed. While the theories may yield the same results as standard QM, they do so at the cost of introducing even more `strangeness,' non-locality being only one concern.
	
	Given the role of the extra $y$ variables, is entropic quantum dynamics compatible with Bell's theorem? We should first note that EQD, like standard quantum mechanics is a non-local theory. The derivation is formulated in configuration space where the dynamics describe the system as a whole, which has immediate non-local consequences. Furthermore, entropic dynamics is not a hidden variable theory---at least not in the usual sense. While the extra $y$ variables do represent some facet of reality that is `hidden' away, precise knowledge of the $y$ variables does not guarantee a deterministic evolution of the system. The system would still jump to some other position in configuration space in a probabilistic way. Whether there is some underlying deterministic theory driving the apparent probabilistic steps of the system is still not known, but the formulation of entropic dynamics remains indifferent.

\section{Conclusions}

	In this chapter we presented a derivation of quantum mechanics from a purely informational approach. It is very important to recognize that entropic dynamics does not discard all of the results of quantum mechanics but simply reinterprets them by introducing more fundamental, informational assumptions. This strategy presents some interesting consequences that will be discussed in the next few chapters.

\mychapter{Information and Symmetry}
	\label{chapter:symmetry}
	The dynamics of a system in entropic dynamics are guided by information. In this scheme, an \emph{observer} is some rational agent capable of using this information to form inferences. Different observers are not guaranteed to have access to the \emph{same} information. A \emph{frame of reference} is characterized by the particular information available to, or rather the state of knowledge of, an observer. In some situations, different observers would come to different conclusions. There are cases, however, where different observers are able to make the same inferences despite having different available information. This equivalence of inferences is called a \emph{symmetry} and is the subject of this chapter.

\section{Transformations}

	From an informational point of view, we call a \emph{transformation} a change in the information used to make inferences. Transformations are typically divided into two distinct classes: active and passive. 

	\subsection{Passive Transformations}
	
		A \emph{passive transformation} is a change in the \emph{description} of a system. Essentially, a passive transformation refers to one system described by two different observers. There is nothing physical here. The actual dynamics of a system are independent of the descriptions given by the observers.
		
\begin{figure}
			\centering
			\subfloat[A passive transformation.]{
				\includegraphics[scale=0.7]{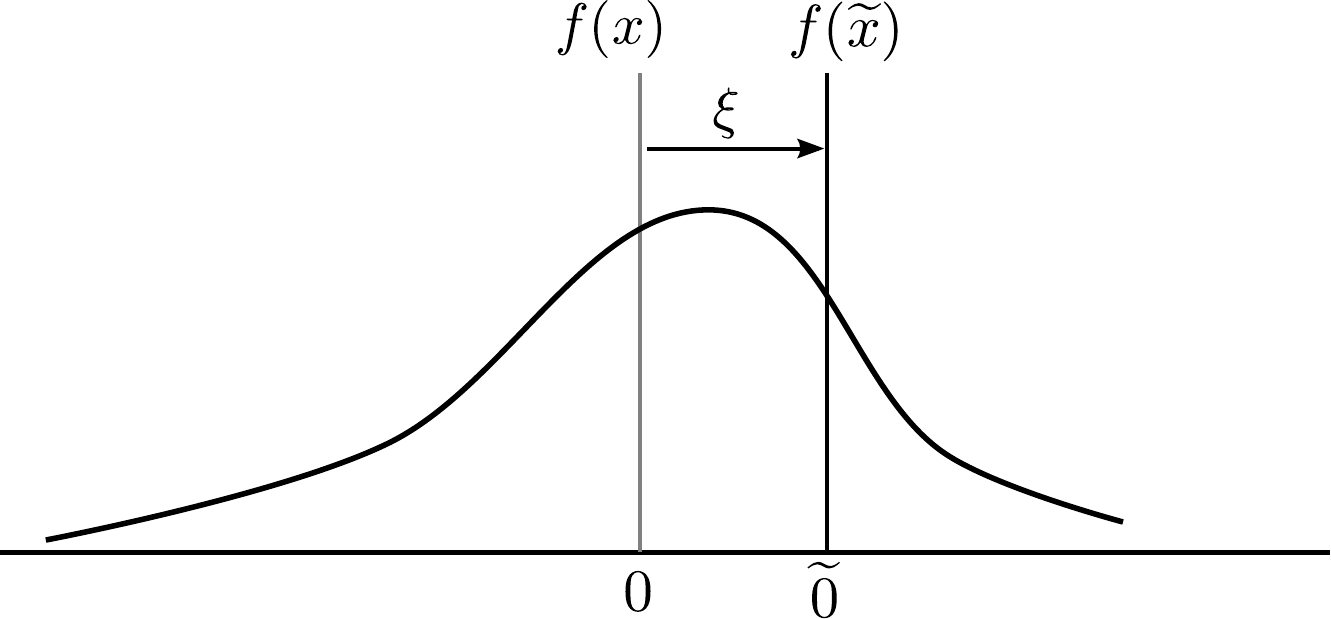}
				\label{fig:passive}
			}
			\\
			\subfloat[An active transformation.]
			{
				\includegraphics[scale=0.7]{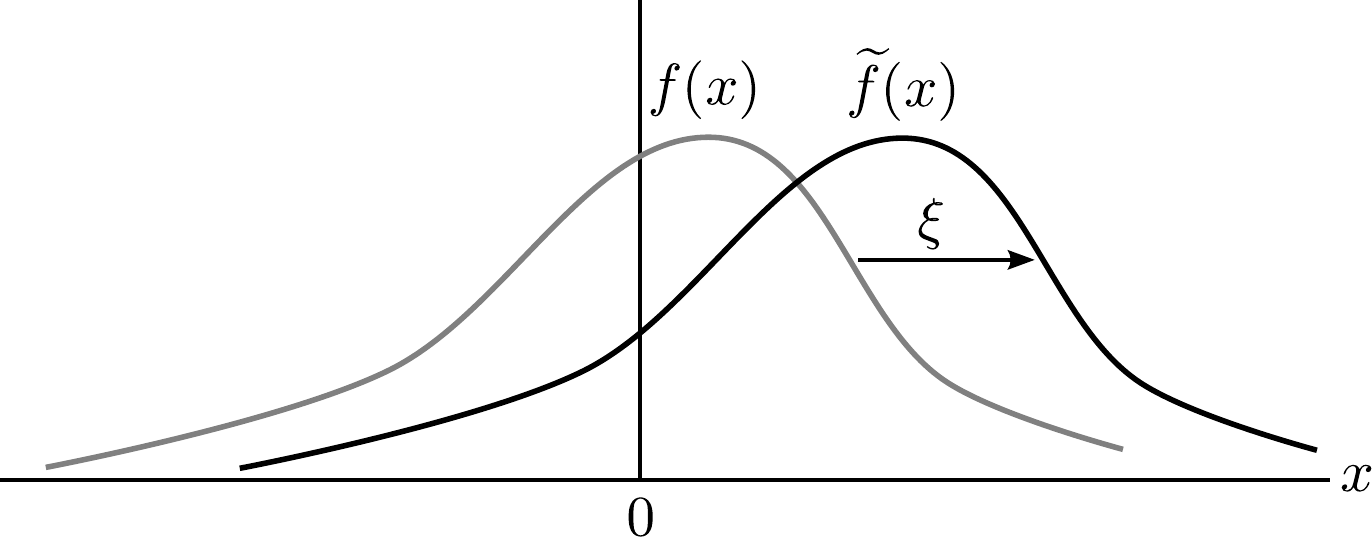}
				\label{fig:active}
			}
			\caption{Passive and active transformations in the case of spatial translation.}
			\label{fig:transformations}
		\end{figure}

		The most common type of passive transformation is a passive coordinate transformation. If the dynamics of a system is described by an observer in the space $\mathcal{X}$, a different observer in a space $\widetilde{\mathcal{X}}$ would describe the same dynamics in a different way. Mathematically, the transformation is a one-to-one mapping $T$ that is, in general, non-linear \cite{houtappel_et_al:1965}. The mapping transforms a point $x$ in the space $\mathcal{X}$ to a point $\tilde{x}$ in $\widetilde{\mathcal{X}}$,
		\begin{equation}
			x \overset{T}{\longrightarrow} \tilde{x} \ .
		\end{equation}
		For example, consider two observers separated by a spatial translation $\xi$,
		\begin{equation}
			\tilde{x} = x + \xi \ .
		\end{equation}
		Translating the coordinates by $\xi$ has the effect of translating a function of those coordinates in the \emph{opposite} direction (cf. Figure \ref{fig:passive}).
		
	\subsection{Active Transformations}
	
		While a passive transformation refers to two different descriptions of the same system by two different observers, an \emph{active transformation} refers to two different states of the same system (or two entirely different systems) described by one observer. An active transformation actively changes the system to a different, transformed system.
			
		Again, a common type of active transformation is an active coordinate transformation. As in the case of passive transformations, the active transformation is a one-to-one mapping $T$. In this case, however, the mapping transforms functions of the coordinates as opposed to the coordinates themselves,
		\begin{equation}
			f(x) \overset{T}{\longrightarrow} \widetilde{f}(x) \ .
		\end{equation}
		The transformation leads to a new function of the same coordinates. Using the example of a spatial translation by $\xi$, the function as actively translated (cf. Figure \ref{fig:active}). In quantum mechanics, this transformation takes the form of a displacement operator acting on the wave function.

\section{Symmetry}

	A \emph{symmetry} is the inability to distinguish physical situations \cite{peres:1993}. More precisely, there is a symmetry between physical situations when the laws of physics and the observations that are correlated by those laws are invariant---at least in some limited way\footnote{As an example, the Galilean transformation exhibits an observable, relative phase difference in coherent superpositions of wave functions with different mass, invalidating the Galilean symmetry \cite{kaempffer:1965}. Legislating a superselection rule prohibiting this kind of superposition in non-relativistic quantum mechanics restores the validity of the symmetry.}. Unobservable quantities, however, are not subject to this requirement. For example, in section \ref{subsec:symmetry_dynamical}, we will show how the phase and potentials in quantum mechanics are free to change (though not arbitrarily).
	
	Informationally, a symmetry implies a notion of an \emph{equivalence of information}. While the information in two physical situations may be substantially different, if the resulting inferences are invariant, the information is said to be equivalent. There are three classes of symmetry that are of interest.

	\subsection{Internal Symmetry}
	
		An \emph{internal symmetry} is a property of an object or a system. It is independent of the space where the object lives and the laws that describe the dynamics of the object. If an object is indistinguishable after an \emph{active} transformation then it has internal symmetry. (If the space in which the object lives exhibits a corresponding geometrical symmetry then this is true for passive transformations as well.) For example, an equilateral triangle is symmetric under rotations of 120$^\circ$ and reflections about its bisectors. Applying any one of these active transformations leaves the laws and predictions unchanged.

	\subsection{Dynamical Symmetry}
		\label{subsec:symmetry_dynamical}
	
		A \emph{dynamical symmetry} is a property of the laws of physics. Here we mean a dynamical symmetry in the way used by Wigner \cite{wigner:1967}. In many mathematical laws there exists the freedom to change quantities in a particular way that leaves the laws unchanged. Since they are a property of the mathematical laws, dynamical symmetries can be found in expressions that have no validity as law of physics. However, when the laws do find application to nature, they may have implications for our description of nature. (For example, the dynamical symmetry of gauge transformations in electrodynamics combined with energy conservation is thought to result in charge conservation \cite{wigner:1949}.)
			
		A trivial dynamical symmetry can be seen in the phase of the wave function. We can always transform the phase of the wave function to a new phase,
		\begin{equation}
			\phi \ \rightarrow \ \tilde{\phi} = \phi + C \ ,
		\end{equation}
		where $C$ is some real constant. This transformation is called a \emph{global gauge transformation}, and it trivially leaves the current velocity invariant,
		\begin{equation}
			\tilde{v}_a  = \partial_a \tilde{\phi} = \partial_a \phi \ .
		\end{equation}
		Additionally, the constant phase shift can be canceled from each term in the linear Schr\"odinger equation (\ref{eq:schrod}), leaving it invariant as well. In entropic dynamics, this constant phase shift can only arise from a constant shift in the entropy of the $y$ variables,
		\begin{equation}
			S \ \rightarrow \ \widetilde{S} = S + C \ .
		\end{equation}
		Here we see two situations with different information about the extra $y$ variables that lead to entropies that differ by a constant. However, despite this different information, the resulting inferences are the same. Such a symmetry is called a \emph{gauge symmetry}.
		
		A non-trivial example of a dynamical symmetry can be seen in section \ref{sec:eqd_external_em}. We saw how introducing an additional constraint on the displacement of a particle in a particular direction introduces an external electromagnetic field \cite{caticha:2010}. The choice of constraint leading to the dynamical equations (i.e.\ the Schr\"odinger equation) is not unique. Consider a transformed external field $\widetilde{A}_a$,
		\begin{equation}
			A_a \ \rightarrow \ \widetilde{A}_a = A_a + \partial_a f \ ,
		\end{equation}
		where $f = f(x,t)$ is some arbitrary function. The field constraint is then
		\begin{equation}
			\langle \Delta x^a \widetilde{A}_a\rangle = C \ .
		\end{equation}
		This new constraint leads to a transformed current velocity
		\begin{equation}
			\tilde{v}_a = \frac{\hbar}{m} \left( \partial_a\widetilde{\phi} - \frac{e}{\hbar c}\widetilde{A}_a \right) \ ,
		\end{equation}
		which is seen to be invariant ($\tilde{v}_a = v_a$) if the phase (and therefore the entropy) transforms as
		\begin{equation}
			\phi \ \rightarrow \ \tilde{\phi} = \phi + \frac{e}{\hbar c}f \ .
		\end{equation}
		Finally, we see that the Schr\"odinger equation is also invariant provided the potential transforms as
		\begin{equation}
			V \ \rightarrow \ \widetilde{V} = V + \frac{e}{c}\,\partial_t f \ .
		\end{equation}
		
		This type of transformation is called a \emph{local gauge transformation} as the phase is changed at each point by the function $f(x,t)$ as opposed to a global change. Once again, we see two informationally different situations leading to identical inferences. In this case, the symmetry is due to an equivalence of information between the constraints imposed in the two different frames of reference.

	\subsection{Geometrical Symmetry}
	
		An important class of symmetry is a \emph{geometrical symmetry}. This is a symmetry of the space where objects reside. We take it for granted that the laws of nature are the same at any point in space or time. This need not be the case. It is a property of reality that the laws of nature are symmetric in this way. In fact, this symmetry can even be expanded to frames moving with constant velocity and even further to accelerating frames, as we will show in the next chapter.
		
		Geometrical symmetries are important because they allow us to repeat inferences. Rarely are the conditions for subsequent experiments precisely the same. Why should we get the same results when the experiment is done at a later time? Or in a different place? In some cases we don't. For particular inferences, however, experiments can be repeated provided all of the \emph{relevant} information is the same.
			
		It should be evident from figure \ref{fig:transformations} that active and passive transformations have inverse effects on a system \cite{peres:1993}. It is invariance under the combination of active and passive transformations that forms a geometrical symmetry. More precisely, there is a geometrical symmetry when
		\begin{equation}
			\widetilde{f}(\tilde{x}) = f(x) \ ,
		\end{equation}
		for all observables and when the laws connecting these observables take the same form.
	
\section{Geometrical Symmetries in Entropic Quantum Dynamics}

	We have already shown an example of a dynamical symmetry in section \ref{subsec:symmetry_dynamical}. In the next chapter we will demonstrate an example of a geometrical symmetry. Before that, however, we should single out an important condition for such a symmetry. If two observers differ by a symmetry transformation, we require that the predictions in the two frames of reference coincide. For any given time $t$ and corresponding $\tilde{t}$, the probabilities assigned to a particular position should be identical. That is,
	\begin{equation}
		\label{eq:sym_con}
		\tilde{\rho}(\tilde{x},\tilde{t})\, d\tilde{x} = \rho(x,t)\, dx \ .
	\end{equation}
	This symmetry condition has implications for the transition probabilities as well. Recalling that the transition probability is related to $\rho(x',t')$ by (\ref{eq:rho_def}), we can replace the probability densities in the previous expression. Then employing the symmetry condition once more gives
	\begin{equation}
		 \int\! dx\ \rho(x,t) \big( P(\tilde{x}'|\tilde{x})\, d\tilde{x}' \big) = \int\! dx\ \rho(x,t) \big( P(x'|x)  dx'\, \big) \ .
	\end{equation}
	If the transformation connecting the two frames is spatially uniform, the requirement that this expression hold for all times implies
	\begin{equation}
		\label{eq:sym_trans}
		P(\tilde{x}'|\tilde{x})\ d\tilde{x}' = P(x'|x) \ dx' \ .
	\end{equation}
	Therefore, when the transformation depends solely on time, the transition probabilities must be invariant as well.

\section{Conclusions}
	The concept of symmetries is extremely useful in physics, and their role in entropic dynamics is no exception. For a transformation relating two observers to qualify as a symmetry, the observers must have \emph{equivalent} states of knowledge. In EQD, this equivalence of information is reflected by the symmetry condition (\ref{eq:sym_con}) and, ultimately, in the invariance of the Schr\"odinger equation. In addition to the insights brought to symmetry by this informational perspective, all of the usual symmetry methods in standard quantum mechanics are still very much valid.

\mychapter{Generalized Galilean Transformations}
	\label{chapter:galilean}
	The standard \emph{Galilean transformation} is a boost to a frame moving with a constant velocity,
\begin{equation}
	\tilde{x}^a = x^a + v_0^{\ a} t \ , \qquad \tilde{t} = t \ ,
\end{equation}
where $v_0^{\ a}$ is a constant. This can be generalized to what is known as the \emph{general Galilean transformation} by including a static rotation and spatial and temporal shifts,
\begin{equation}
	\tilde{x}^a = R^a_{\ b} x^b + v_0^{\ a} t +  x_0^{\ a}\ , \qquad \tilde{t} = t + t_0\ ,
\end{equation}
with $x_0^{\ a}$ and $t_0$ both constant \cite{kaempffer:1965}. One can also generalize the boost to frames with an arbitrary, rigid acceleration. This \emph{extended Galilean transformation} (EGT) is given by
\begin{equation}
	\tilde{x}^a = x^a + \xi^a(t) \ , \qquad \tilde{t} = t \ ,
\end{equation}
where $\xi(t)$ is an arbitrary function of time \cite{rosen:1972}.

The extended Galilean transformation is of particular interest because it retains residual features of both special and general relativity in non-relativistic quantum mechanics \cite{rosen:1972,greenberger:1979,greenberger:2001}. These novel features of the transformation are due solely to the velocity difference between the two frames---the rotation and temporal shift are of little interest. In this chapter, we follow our work in  \cite{johnson:2010} and demonstrate how this extended Galilean transformation results in a geometrical symmetry through an equivalence of information. 

\section{The Transformed Frame}
	Consider a new observer describing the dynamics of a particle. In the new observer's frame of reference, the particle lives in a 3-dimensional space $\smash{\widetilde{\mathcal{X}}}$. We are not assuming a definition of time yet, only that the particle will take small steps. Accordingly, the transformation connecting the two spaces is $\tilde{x}^a = x^a + \xi^a(\theta)$, where $\xi$ is an arbitrary displacement and $\theta$ is a parameter that is free to vary as the particle takes steps. (Inclusion of a static rotation is straightforward.) Both the metric and the volume element $d^3x$ are invariant as $\xi$ is spatially uniform.

	The introduction of dynamics in the transformed frame follows a very close parallel with the original derivation. Applying the method of maximum entropy subject to normalization and the constraint of small steps leads to a transformed transition probability $\smash{\widetilde{P}}(\tilde{x}'|\tilde{x})$ of the same form as (\ref{eq:raw_prob}) but with a new Lagrange multiplier $\tilde{\alpha}(\tilde{x})$ and a transformed entropy $\smash{\widetilde{S}}(\tilde{x})$. The transformed entropy represents the new observer's state of knowledge about the $\tilde{y}$ variables.
	
\section{Time}

	We wish to model the very same Newtonian time in $\smash{\widetilde{\mathcal{X}}}$ so that time flows not only at the same rate everywhere in space but at the same rate in every frame. Hence we define $\alpha(x,t) = \tilde{\alpha}(\tilde{x},\tilde{t}) = \tau/\Delta{\tilde{t}} = $ constant. Now we can identify the parameter $\theta$ with the time $\tilde{t}$ and specify the full transformation
	\begin{equation}
		\label{eq:trans}
		\tilde{x}^a = x^a + \xi^a(t) \ , \qquad \tilde{t} = t \ ,
	\end{equation}
	with derivatives,
	\begin{equation}
		\tilde{\partial}_t = \partial_t - \dot{\xi}^a \partial_a \ \qquad \text{and} \qquad \tilde{\partial}_a = \partial_a \ ,
	\end{equation}
	where $\tilde{\partial}_a = \partial/\partial \tilde{x}^a$ and $\dot{\xi}^a = \partial_t \xi^a(t)$.

	Upon substituting the definition of $\tilde{\alpha}$ into the transition probability and expanding about the maximum displacement $\Delta\tilde{\bar{x}}$,
	\begin{equation}
		\label{eq:transformed_approx_trans_prob}
		\widetilde{P}(\tilde{x}'|\tilde{x}) \approx \frac{1}{\widetilde{Z}} \exp \left[ -\frac{m}{2\hbar \Delta \tilde{t}}\, \delta_{ab} (\Delta \tilde{x}^a - \Delta \tilde{\bar{x}}^a)(\Delta \tilde{x}^b - \Delta \tilde{\bar{x}}^b) \right] \ ,
	\end{equation}
	where the displacement $\Delta \tilde{x}^a = \tilde{b}^a(\tilde{x})\Delta \tilde{t} + \Delta \tilde{w}^a$. The drift velocity $\tilde{b}^a(\tilde{x})$ is given by
	\begin{equation}
		\label{eq:trans_future_drift}
		\langle \Delta \tilde{x}^a \rangle = \Delta\tilde{\bar{x}} = \tilde{b}^a(\tilde{x})\Delta \tilde{t} \quad \text{where} \quad \tilde{b}^a(\tilde{x}) = \frac{\hbar}{m} \tilde{\partial}^a \widetilde{S}(\tilde{x}) \ ,
	\end{equation}
	while the fluctuations are unchanged by the transformation.

\section{Enforcing Symmetry}

	We demand that the dynamics obey the symmetry condition (\ref{eq:sym_con}). Since the extended Galilean transformation depends only on time, the condition also asserts that the transition probabilities are invariant (\ref{eq:sym_trans}). Comparing these probabilities in both frames (\ref{eq:approx_trans_prob}, \ref{eq:transformed_approx_trans_prob}) implies that their exponents must be equal, as the normalization function $\smash{\widetilde{Z}} = Z$. This implies
	\begin{equation}
		\label{eq:sym1}
		\Delta \tilde{x}^a - \Delta \tilde{\bar{x}}^a = \pm (\Delta x^a - \Delta \bar{x}^a) \ .
	\end{equation}
	The negative sign is rejected as being inconsistent with the limit of $\xi \rightarrow 0$, where $\tilde{x} \rightarrow x$. The transformed displacement can be expressed in terms of the original displacement, $\Delta \tilde{x}^a = \Delta x^a + \Delta \xi^a$. By substituting this and the expressions for the mean displacements in both frames (\ref{eq:future_drift}, \ref{eq:trans_future_drift}) and rearranging, we have
	\begin{equation}
		\tilde{\partial}_a (\Delta S) = \frac{m}{\hbar}\frac{\Delta \xi}{\Delta t} \ ,
	\end{equation}
	where
	\begin{equation}
		\Delta S = \widetilde{S}(\tilde{x}) - S(x)
	\end{equation}
	is a shift in the entropy. Solving the differential equation for the entropy shift and taking the limit $\Delta t \rightarrow 0$ gives
	\begin{equation}
		\label{eq:entropy_shift_mostly}
		\Delta S = \frac{m}{\hbar} \left( \dot{\xi}^a \tilde{x}_a +c(t) \right) \ ,
	\end{equation}
	where $c(t)$ is a constant of integration.

	The transformed drift and osmotic velocities relative to the original frame are then straightforward to calculate,
	\begin{equation}
		\label{eq:trans_velocities}
		\tilde{b}^a = b^a + \dot{\xi}^a \qquad \text{and} \qquad \tilde{u}^a = - \frac{\hbar}{2m} \tilde{\partial}^a \log \tilde{\rho} = u^a \ .
	\end{equation}
	The drift velocity follows from (\ref{eq:trans_future_drift}) and (\ref{eq:entropy_shift_mostly}). The invariance of the osmotic velocity follows directly from the symmetry condition (\ref{eq:sym_con}). If we let $\tilde{v}^a = \tilde{b}^a + \tilde{u}^a$ be the transformed current velocity, the transformed FP equation is covariant,
	\begin{equation}
		\label{eq:trans_fp}
		\tilde{\partial}_t \tilde{\rho} = -\tilde{\partial}_a(\tilde{v}^a\tilde{\rho}) \ .
	\end{equation}
	Additionally, it follows from (\ref{eq:trans_velocities}) that the current velocity is also shifted by the same amount as the drift velocities,
	\begin{equation}
		\label{eq:current_velocity_shift}
		\tilde{v}^a = v^a + \dot{\xi}^a \ . 
	\end{equation}
	Finally, we can express the current velocity as a gradient,
	\begin{equation}
		\label{eq:transformed_velocity_grad}
		\tilde{v}^a = \frac{\hbar}{m} \tilde{\partial}^a \tilde{\phi} \ ,
	\end{equation}
	where
	\begin{equation}
		\label{eq:tranformed_phase}
		\tilde{\phi}(\tilde{x},\tilde{t}) = \phi(x,t) + \frac{m}{\hbar} \left( \dot{\xi}^a \tilde{x}_a +c(t) \right) \ ,
	\end{equation}
	showing that the transformation causes a phase shift, as expected.

\section{The Schr\"odinger Equation}

	We must now introduce a transformed energy functional so as to allow the entropy to participate in the dynamics, but we cannot assume that either the energy functional or the conservation condition take the same form as in the original frame (\ref{eq:energy1}, \ref{eq:energy_cons}). Rather, we start with the original conservation of energy condition (\ref{eq:energy_cons}) and energy functional (\ref{eq:energy1}). Upon expressing the current velocity in terms of the transformed coordinates ($\smash{v^2 = \tilde{v}^2 - 2\dot{\xi}^a\tilde{v}_a + \dot{\xi}^2}$) and simplifying,
	\begin{eqnarray}
		\label{eq:trans_energy2}
		\dot{E} - \int \! dx\ \rho \partial_t V &=& \int \! dx \ \bigg[ \partial_t \rho \left(\tfrac{1}{2}m \tilde{v}^2 + \tfrac{1}{2}m u^2 + V -m\dot{\xi}^a\tilde{v}_a \right) \nonumber \\
		&&\quad+\ \rho\left( \tfrac{1}{2}m\partial_t u^2 + mv_a\partial_t \tilde{v}^a \right)\bigg] - \int \! dx\ \rho m\ddot{\xi}^a v_a  = 0 \ .
	\end{eqnarray}
	
	The $\ddot{\xi}$ integral in (\ref{eq:trans_energy2}) arises from the velocity cross term and is particularly interesting. Inserting the $\mathcal{X}$ current velocity (\ref{eq:current_v_grad_phi}), integrating by parts, and substituting the original FP equation (\ref{eq:fp}) results in
	\begin{eqnarray}
		\label{eq:grav_integral}
		\int \! dx\ (\hbar\ddot{\xi}^a) (\rho \partial_a\phi) &=& -\int \! dx \ \hbar\ddot{\xi}^a(x_a +d_a(t)) \ \partial^b(\rho \partial_b\phi) \nonumber \\
		&=& \int \! dx \ \partial_t \rho \, m \ddot{\xi}^a\left(\tilde{x}_a + d_a(t)\right) \ ,
	\end{eqnarray}
	where $d^a(t)$ is a constant of integration. As we will show later, its arbitrariness reflects the freedom in choosing the zero of the effective gravitational potential introduced by the accelerating frame, which is a global gauge transformation. This is not unique to the EQD version of this transformation but exists in the standard formulation as well.
	
	The remaining terms in (\ref{eq:trans_energy2}) are handled in the same way as the energy condition in $\mathcal{X}$ (\ref{eq:energy1}). Requiring the condition to hold for arbitrary choices of $\tilde{\rho}$ and $\tilde{\phi}$ and manipulating with integration by parts and the FP equations results in a transformed phase equation,
	\begin{equation}
		\label{eq:trans_pde_phi}
		\hbar\tilde{\partial}_t \tilde{\phi} + \frac{\hbar^2}{2m} (\tilde{\partial}_a \tilde{\phi})^2 + \widetilde{V} -\frac{\hbar^2}{2 m} \frac{\tilde{\nabla}^2 \tilde{\rho}^{1/2}}{\tilde{\rho}^{1/2}} = 0 \ ,
	\end{equation}
	provided the transformed potential energy is
	\begin{equation}
		\label{eq:full_grav_pot}
		\widetilde{V} = V - m \ddot{\xi}^a \left(\tilde{x}_a + d_a(t)\right) \ .
	\end{equation}
	By combining the new phase equation with the transformed FP equation (\ref{eq:trans_fp}) into a complex function $\smash{\widetilde{\Psi} = \tilde{\rho}^{1/2} e^{i\tilde{\phi}}}$, we obtain the Schr\"odinger equation in terms of the transformed quantities,
	\begin{equation}
		\label{eq:trans:schrod}
		i\hbar \frac{\partial\widetilde{\Psi}}{\partial \tilde{t}} = -\frac{\hbar^2}{2m}\tilde{\nabla}^2\widetilde{\Psi} + \widetilde{V}\widetilde{\Psi} \ .
	\end{equation}
	
	To determine the integration constant $c(t)$ in the entropy shift (\ref{eq:entropy_shift_mostly}), we simply express the phase equation (\ref{eq:trans_pde_phi}) in the original coordinates. This results in a differential equation with solution
	\begin{equation}
		\label{eq:c_full}
		c(t) = -\frac{1}{2} \int \! dt \ \left(\dot{\xi}^2 - 2\ddot{\xi}^a d_a(t) \right) \ .
	\end{equation}
	Since the choice of gauge is arbitrary and one can always choose a new $\smash{\widetilde{\Psi}}$ differing by a phase so as to eliminate $d^a(t)$, we set it to $0$. This choice is standard and simplifies the form of the phase shift (\ref{eq:c_full}) and the potential (\ref{eq:full_grav_pot}),
	\begin{equation}
		c(t) = -\frac{1}{2} \int \! dt \ \dot{\xi}^2 \ , \qquad \widetilde{V} = V - m \ddot{\xi}^a \tilde{x}_a \ .
	\end{equation}

	Up to a gauge freedom, we have determined the difference between the entropies of the observers,
	\begin{equation}
		\label{eq:entropy_diff_final}
		\Delta S = \widetilde{S}(\tilde{x},\tilde{t}) - S(x,t) = \frac{m}{\hbar} \left( \dot{\xi}^a \tilde{x}_a - \frac{1}{2} \int dt \ \dot{\xi}^2\right) \ .
	\end{equation}
	Again, this difference is also the phase shift between the frames, and it is the same result that the EGT yields in the standard formulation of QM. In entropic dynamics, however, it takes a new meaning as the relation between the states of knowledge which two observers \emph{must} have in order for the extended Galilean transformation to qualify as a symmetry.

	The transformed dynamical equation (\ref{eq:trans_pde_phi}) has the same form as the original (\ref{eq:pde_phi}). One could also obtain this same result by starting with a transformed energy functional,
	\begin{equation}
		\label{eq:trans_energy_final}
		\widetilde{E}[\tilde{\rho},\widetilde{S}] = \int \! d^3\tilde{x} \ \tilde{\rho}(\tilde{x},\tilde{t}) \left( \tfrac{1}{2}m\tilde{v}^2 + \tfrac{1}{2}m \tilde{u}^2 + \widetilde{V}(\tilde{x},\tilde{t}) \right) \ ,
	\end{equation}
	and conservation condition,
	\begin{equation}
		\frac{d}{d\tilde{t}}\, \widetilde{E} = \int \! d\tilde{x} \ \tilde{\partial}_t\widetilde{V} \ .
	\end{equation}
	This implies that although the energy and its rate of change in the two frames is very different (as illustrated by the transformed potential (\ref{eq:full_grav_pot})) the energy functional does indeed take the same form and the conservation requirement is covariant.

\section{Special Relativity}
	As we showed in equations (\ref{eq:current_velocity_shift}--\ref{eq:tranformed_phase}), the first term in the entropy shift (\ref{eq:entropy_diff_final}) is required for the momentum to transform properly. The second term, however, tells a more interesting story. Dividing by $c^2$ and rearranging, the integral can be written as
	\begin{eqnarray}
		\frac{1}{2c^2} \int \! dt \ \dot{\xi}^2 = t - \int \! dt \ \left( 1 - \frac{\dot{\xi}^2}{2c^2} \right) \approx t - \int \! dt \ \left(1 - \frac{\dot{\xi}^2}{c^2} \right)^{1/2} \ .
	\end{eqnarray}
	The integral in the second step is a first-order approximation of the proper time of the moving observer \cite{greenberger:1979}. We see that the second term of the entropy shift is a residue of special relativity due to the difference in proper time between the two frames of reference. (This is not a mathematical artifact but is a real, observable phase shift.) Although our non-relativistic formulation of quantum mechanics makes no distinction between coordinate and proper time, this residual effect indicates that a relativistic quantum entropic theory would necessarily include different definitions of time in order to properly reflect the observers' differing states of information. This is, however, a subject for future work.

\section{Information and the Strong Equivalence Principle} 

	In the special case where $\dot{\xi}^a = v_0^{\ a}$ is a constant velocity, the EGT reduces to the standard Galilean transformation with entropy shift and potential given by
	\begin{equation}
		\Delta S = \frac{m}{\hbar} \left( v_0^{\ a} \tilde{x}_a - \frac{1}{2} v_0^{\ 2}t \right) \qquad \text{and} \qquad \Delta V = 0 \ .
	\end{equation}
	And in the case of a constant acceleration, $\ddot{\xi}^a = g^a$, the entropy shift and potential are
	\begin{equation}
		\label{eq:special_uniform_a}
		\Delta S = \frac{m}{\hbar} \left( g^a \tilde{x}_at - \frac{1}{6} g^2t^3 \right) \qquad \text{and} \qquad \Delta V = - m g^a \tilde{x}_a \ .
	\end{equation}
	The transformed potential has an additional term that is of the same form as a uniform gravitational field.

	The additional term in the potential (\ref{eq:full_grav_pot}) is an effective gravitational potential arising from the acceleration of the $\smash{\widetilde{\mathcal{X}}}$ frame. The standard interpretation of this potential is that it enters via the strong equivalence principle \cite{rosen:1972}. The strong equivalence principle from general relativity states that gravitational effects are equivalent to the fictitious effects that arise in non-inertial frames. We can see this in the special case of a constantly accelerating frame (\ref{eq:special_uniform_a}), where this `fictitious' potential shift is \emph{indistinguishable} from a uniform gravitational field.

	In EQD, however, the general covariance implied by the strong equivalence principle is the result of an \emph{equivalence of information}. Just as we discussed for the electromagnetic gauge transformation in section \ref{subsec:symmetry_dynamical}, we have two situations with substantially different information that lead to the very same inferences. Perhaps the strong equivalence principle introduced by Einstein in his theory of general relativity is due to this very equivalence of information. This result is potentially an opening to an entropic explanation of gravity.

\mychapter{The Measurement Problem}
	\label{chapter:measurement}
	The subject of measurement is perhaps the most hotly debated problem in quantum theory \cite{reece:1973}. The problem of measurement sits at the interface between the strange, probabilistic quantum world and the deterministic classical one.

The goal of this chapter is to lay out a theory of measurement in entropic dynamics. We begin by reviewing some mathematical formalism that is common to both EQD and the standard QM approach. Then in section \ref{sec:measurement_problem} we discuss what is known as ``the measurement problem'' in quantum mechanics. This section illustrates the difficulties in the standard treatment of measurement, which has lead to countless new theories and interpretations.

In entropic dynamics, position is the sole observable. We will show how a theory built only on position can account for the vast array of measurements one can perform on a quantum system. In the standard quantum theory, the Born rule is a postulate that determines the probabilities of outcomes of a measurement. However, our informational approach in EQD leaves no room for any measurement postulates. Fortunately, the Born rule need not be postulated. For position, the rule is a direct consequence of our statistical model. For measurements of other variables, we show that the Born rule is a natural consequence of the unitary evolution of the Schr\"odinger equation.

Another postulate that must be examined is the projection postulate---that after interacting with a measuring device the wave function must be left in an eigenstate of the operator representing the device. We discuss how this postulate originates when one forces a realistic interpretation on the wave function. It is reinforced by the over-application of a very specialized experimental procedure known as filtering. We show how in this special case the ME method can be used to update the wave function when new, relevant information is available. Such updating is only possible in entropic dynamics because the entire wave function (including the phase) is statistical in nature.

We conclude the chapter with discussions of a number of topics relevant to measurement. We show how the classical determinism of macroscopic objects arises in a quantum system with many particles. Then we discuss the interpretation of the uncertainty principle in entropic dynamics. Finally, we include a brief comment on sequential measurements. In stochastic mechanics, the multitime correlations in such measurements were very troublesome. In entropic dynamics they are handled quite easily.

\section{Mathematical Preliminaries}

	In this section we provide some mathematical preliminaries that are common to both entropic dynamics and the standard formulation of quantum theory \cite{ballentine:1998,sakurai:1994}.
	
	The linearity of the Schr\"odinger equation (SE) presents us with numerous mathematical advantages. Linear, homogeneous differential equations like the SE obey the \emph{superposition principle}. Consider two wave functions $\Psi_1(x,t)$ and $\Psi_2(x,t)$. If each is a solution to the SE, then a linear superposition of the two functions must also be a solution. A general superposition of the two wave functions is
	\begin{equation}
		\Psi(x,t) = c_1 \Psi_1(x,t) + c_2 \Psi_2(x,t) \ ,
	\end{equation}
	where $c_1$ and $c_2$ are complex constants. One can verify that the superposition is a solution by simple substitution. The generalization to a superposition of an arbitrary number of wave functions is immediate. Note that the principle of superposition is a \emph{mathematical} property of the Schr\"odinger equation.

	This linearity of the SE allows us to treat states as vectors in a linear vector space with the complex coefficients acting as the components of the vectors. A set of vectors $\phi_i$ is said to be \emph{linearly independent} if no vector in the set can be expressed as a superposition of any of the others. A set of linearly independent vectors forms a \emph{basis} if the vectors span the space. This implies that any vector in the space may be represented as a linear superposition of the basis vectors.

	The \emph{inner product} of two vectors is a binary operation that assigns a scalar for each pair of vectors. We define the inner product of two vectors as
	\begin{equation}
		(\psi, \phi) = \int_{-\infty}^\infty\!dx\ \psi^*(x) \phi(x) \ .
	\end{equation}
	The inner product of a vector with itself is the square of the norm of the vector,
	\begin{equation}
		(\phi, \phi) = \int_{-\infty}^\infty\!dx\ \phi^*(x) \phi(x) = ||\phi||^2 \ .
	\end{equation}
	A complete vector space equipped with an inner product is a \emph{Hilbert space}. In order to interpret the vectors in the Hilbert space as probability densities, they must first be normalized so that
	\begin{equation}
		\Psi^*(x,t) \Psi(x,t) = \rho(x,t) \qquad \text{with} \qquad \int\!dx\ \rho(x,t) = 1 \ .
	\end{equation}
	To simplify notation, we will only consider vectors normalized in this way.

	A set of normalized vectors $\{\phi_i\} $ is called \emph{orthonormal} if the vectors are all mutually orthogonal. This means that the inner products $(\phi_i,\phi_j) = \delta_{ij}$, for all $i,j$.

	\subsection{Bra-ket Notation}

		As this point it is convenient to introduce a concise notation developed by Dirac called \emph{bra-ket notation}. In this scheme we write the vectors as \emph{kets}, $|\phi\rangle$. For each ket, there is a corresponding vector in a dual vector space called a \emph{bra}, $\langle \phi|$. A bra combined with a ket implies an inner product,
		\begin{equation}
			\langle \phi|\psi\rangle = (\phi,\psi) \ .
		\end{equation}
		The bras and kets can be labeled by whatever notation is convenient. The relationship between a ket and the corresponding bra is
		\begin{equation}
			|\psi\rangle = c_1|\psi_1\rangle + c_2|\psi_2\rangle \quad \leftrightarrow \quad \langle \psi| = c_1^*\langle\psi_1| + c_2^*\langle\psi_2| \ ,
		\end{equation}
		where $c^*$ indicates the complex conjugate.
		
		If a set of basis vectors $\{ \phi_i\}$ is complete, we can expand any state $|\psi\rangle$ as a superposition of the basis vectors. In bra-ket notation, this is written
		\begin{equation}
			\label{eq:expansion}
			|\psi\rangle = \sum_i c_i|\phi_i\rangle \ ,
		\end{equation}
		where the $c_i$ are complex expansion coefficients. If the eigenstate $|\psi\rangle$ is properly normalized,
		\begin{equation}
			\langle \psi | \psi\rangle = 1 \ ,
		\end{equation}
		which implies that
		\begin{equation}
			\sum_i |c_i|^2 = 1 \ ,
		\end{equation}
		where $|c_i|^2 = c_i^*c_i^{}$.

		The orthonormality of the basis vectors implies that the coefficients are given by $c_i = \langle \phi_i|\psi\rangle$. Substituting the expression for the coefficients into the expansion (\ref{eq:expansion}) results in the \emph{completeness relation},
		\begin{equation}
			\label{eq:discrete_completeness}
			\sum_i |\phi_i\rangle \langle \phi_i| = I \ ,
		\end{equation}
		where $I$ is the identity matrix. 

	\subsection{Linear Operators}
	
		An \emph{operator} is a transformation that maps vectors in one vector space to vectors in another. If $\hat{A}$ is an operator, the result of the operator acting on some vector $|\phi\rangle$ is a new vector, $\hat{A}|\phi\rangle = |\psi\rangle$. An eigenstate of an operator is a vector that satisfies an eigenvalue equation,
		\begin{equation}
			\hat{A}|\phi\rangle = a|\phi\rangle \ ,
		\end{equation}
		where $a$ is a constant. It is common practice to label an eigenstate of an operator by its eigenvalue, $|a\rangle$. A \emph{linear operator} is one that satisfies the following relation,
		\begin{equation}
			\hat{A}\big(c_1|\psi_1\rangle + c_2|\psi_2\rangle\big) = c_1\hat{A}|\psi_1\rangle + c_2\hat{A}|\psi_2\rangle \ .
		\end{equation}
		We will only consider operators that are linear. Operators are associative but not, in general, commutative.

		The \emph{adjoint} (or \emph{Hermitian conjugate}) of an operator is the transpose conjugate of the matrix representing the operator. The adjoint of an operator $\hat{A}$ is written $\hat{A}^\dagger$. The adjoint of a sum of operators is
		\begin{equation}
			(\hat{A} + \hat{B})^\dagger = \hat{A}^\dagger + \hat{B}^\dagger \ ,
		\end{equation}
		while the adjoint of a product of operators reverses the order of the operators,
		\begin{equation}
			(\hat{A}\hat{B})^\dagger = \hat{B}^\dagger\hat{A}^\dagger \ .
		\end{equation}
		An operator that satisfies $\hat{A}^\dagger = \hat{A}$ is said to be \emph{Hermitian} or \emph{self-adjoint}. The eigenstates of a Hermitian operator corresponding to different eigenvalues are orthogonal,
		\begin{equation}
			\langle a_i | a_j\rangle = \delta_{ij} \ .
		\end{equation}
		For a continuous spectrum of eigenvalues, this orthogonality is written
		\begin{equation}
			\label{eq:continuous_ortho}
			\langle a | a'\rangle = \delta(a - a') \ ,
		\end{equation}
		where $\delta(a - a')$ is a Dirac delta function.

		The connection between wave functions expressed in bra-ket notation and those in function form is determined by the eigenstates of the position operator $\hat{x}|x\rangle = x|x\rangle$. From the completeness relation (\ref{eq:discrete_completeness}) and the definition of the inner product,
		\begin{equation}
			\langle \Psi|\Psi\rangle = \int\!dx\ \langle\Psi|x\rangle\langle x| \Psi\rangle = \int\!dx\ \Psi^*(x)\Psi(x) \ .
		\end{equation}
		This implies that the wave function $\Psi(x) = \langle x|\Psi\rangle$ and that the probability density is $\rho(x) = |\langle x|\Psi\rangle|^2$, which is the \emph{Born rule} for \emph{position}.

	\subsection{Unitary Evolution}

		The evolution of the wave function is \emph{unitary}. Consider the evolution of a wave function from an initial time to some later time $t$. We can write the relationship between the initial and final wave functions as
		\begin{equation}
			\label{eq:evolution_rel}
			|\Psi(t)\rangle = \hat{U}(t)|\Psi(0)\rangle \ ,
		\end{equation}
		where $\hat{U}(t)$ is some time-dependent operator. If we substitute $t=0$, this equality requires that $\hat{U}(0) = I$, the identity matrix.
		
		The Schr\"odinger equation (\ref{eq:schrod}) can be written in operator form as,
		\begin{equation}
			i\hbar \frac{\partial}{\partial t}|\Psi(t)\rangle = \hat{H}|\Psi(t)\rangle \ .
		\end{equation}
		If we substitute the evolution relationship (\ref{eq:evolution_rel}) into the Schr\"odinger equation, we get a differential equation for the evolution operator $\hat{U}$,
		\begin{equation}
			i\hbar \frac{\partial \hat{U}}{\partial t} = \hat{H}\hat{U} \ ,
		\end{equation}
		with solution
		\begin{equation}
			\hat{U}(t) = \exp\left[ -\frac{i}{\hbar} \int^t_0\!dt'\ \hat{H}(t') \right] \ .
		\end{equation}
		The Hamiltonian operator is Hermitian (i.e.\ $\hat{H}^\dagger = \hat{H}$), which implies that the evolution operator is unitary, $\hat{U}^\dagger \hat{U} = I$.

		The unitarity of the Schr\"odinger equation has important consequences. This unitarity preserves the inner product,
		\begin{equation}
			\langle \Psi(t) | \Psi(t) \rangle = \langle \Psi(0) | \hat{U}^\dagger \hat{U} | \Psi(0) \rangle = \langle \Psi(0) | \Psi(0) \rangle = 1 \ .
		\end{equation}
		This preservation of the inner product also implies that the orthogonality of the basis states must be preserved as they evolve.

\section{The Measurement Problem}
	\label{sec:measurement_problem}
	In the standard approach to quantum mechanics, the wave function is said to \emph{completely} describe the quantum system \cite{ballentine:1970}. When pushed further, this viewpoint presents numerous conceptual difficulties.
	
	Consider a quantum system that is in a superposition of two eigenstates of an operator $\hat{A}$,
	\begin{equation}
		|\Psi\rangle = c_1 | a_1 \rangle + c_2|a_2\rangle \ .
	\end{equation}
	In the standard approach, the \emph{Born rule} is a postulate that states that the probability of observing the system to be in the state $|a_i\rangle$ is given by the coefficients of the eigenstates, $p_i = |c_i|^2$.
	
	If the wave function is describing some element of reality, measuring that the particle is in some state $|a_i\rangle$ means that wave function must be in the state $|a_i\rangle$ immediately after the measurement \cite{ballentine:1990}. The precise way in which the full wave function $|\Psi\rangle$ evolves into one of the eigenstates is at the heart of the quantum measurement problem. There is no way for the linear Schr\"odinger equation to explain the evolution of the wave function $|\Psi\rangle$ into one of the eigenstates \cite{komar:1962}. This has prompted countless extensions to quantum mechanics that attempt to introduce non-linearity to the evolution. However, the orthodox approach simply postulates a ``collapse'' of the wave function---the \emph{projection postulate}. The interaction with a measurement device is said to somehow collapse the wave function into an eigenstate with probability given by the Born rule. This abrupt, probabilistic change in the wave function stands in stark contrast to the continuous, deterministic evolution specified by the Schr\"odinger equation. Furthermore, it has been shown that the projection postulate has limited utility and even leads to incorrect results \cite{ballentine:1990}.
	
	The problem is exacerbated by considering the inclusion of the measuring apparatus into the wave function. Classical systems like a measurement device should, in principle, be able to be treated as quantum systems with many degrees of freedom. Consider the same system coupled to a measurement device. The combined wave function is
	\begin{equation}
		\label{eq:meas_prob_superpos}
		|\Psi\rangle = c_1 | a_1 \rangle |0\rangle + c_2|a_2\rangle |0\rangle \ ,
	\end{equation}
	where $|0\rangle$ indicates the initial state of the apparatus before measurement. After interacting with the measurement device, the entangled wave function evolves to
	\begin{equation}
		|\Psi'\rangle = \hat{U}|\Psi\rangle = c_1 | a_1 \rangle |\alpha_1\rangle + c_2|a_2\rangle |\alpha_2\rangle \ .
	\end{equation}
	where $\alpha_i$ indicate the pointer states of the apparatus that correspond to a measurement of $a_i$. We now have a wave function that is in a superposition of macroscopic states. In the orthodox approach, such a wave function that ``fully'' describes the system is strange indeed. What does a superposition of classical states physically ``look like,'' and why don't we observe such superpositions?
	
	This bizarre situation was originally pointed out by Schr\"odinger himself in 1935 with his now famous thought experiment involving a cat trapped in a poison chamber \cite{schrodinger:1935} (for English translation see \cite{trimmer:1980}). The hapless feline finds itself in the awkward position of somehow being both alive \emph{and} dead at the same time. Some authors go to the other extreme by denying that the ``liveness'' of the cat is defined until a measurement has been performed---the cat is \emph{neither} live nor dead \emph{until} measured \cite{peres:1988,peres:1984}.
	
	The following sections of this chapter will detail how the entropic interpretation of quantum theory avoids the difficulties in the orthodox approach.

\section{Measurement in EQD}

	The goal of entropic quantum dynamics is to predict one thing---the position of a particle or system of particles specified by the probability density $\rho(x) = \Psi^*(x)\Psi(x)$. The wave function only represents the information available to predict the position of a particle.
	
	Feynman and many other physicists have noted that every measurement appears to be a measurement of position, however indirect it may be. For example, momentum can be measured by measuring the position of a particle after interacting with a known magnetic field. Spin can be measured by measuring the position of a particle after traveling through the inhomogeneous magnetic field of a Stern-Gerlach device.
	
	It may very well be that there is only one true observable in reality: position. In entropic dynamics, however, this reliance on position is built directly into the statistical model. In this section we seek to show how a theory built purely on observation of position can account for the apparently vast array of ``observables.'' In the process, we see that the Born rule need not be postulated for these other observables, but it is a natural consequence of the unitary evolution of the wave function.
	
	The experimental process is divided into three distinct parts. First, a system is prepared in a reproducible way. Then the system is subjected to a measurement that results in a measured position of the system \cite{ballentine:1970}. Finally, the measurement is amplified by a classical system.

	\subsection{State Preparation}
	
		Before a state may be measured, it must be prepared. The state of a quantum system is determined by some reproducible preparation procedure. The determination of the wave function produced by a given preparation procedure amounts to knowing the relevant information about about a system. There are means of preparing states in a systematic way \cite{ballentine:1998}, but the wave function corresponding to a given preparation is typically determined by \emph{calibrating} the device. By performing a number of different measurements, one can infer what the prepared wave function was \cite{stapp:1972}.
	
	\subsection{Measurement}

		A \emph{measurement} consists of subjecting a quantum system to some potential and examining the resulting probability density. The resulting wave functions are simply calculated by the Schr\"odinger equation, which unitarily evolves the initial state to a final state,
		\begin{equation}
			|\Psi(x,t_0)\rangle \rightarrow |\Psi(x,t)\rangle = \hat{U}|\Psi(x,t_0)\rangle \ .
		\end{equation}
		To simplify notation, we will write this expression as 
		\begin{equation}
			|\Psi\rangle \rightarrow |\Psi'\rangle = \hat{U}|\Psi\rangle \ .
		\end{equation}
		
		For simplicity, we will initially consider measurements that have only a discrete set of possible outcome positions. In this case, the continuous probabilities become discrete,
		\begin{equation}
			\rho(x)\, dx = |\langle x|\Psi\rangle|^2\, dx \quad \rightarrow \quad p_i = |\langle x_i|\Psi\rangle|^2 \ .
		\end{equation}
		Consider a eigenstate of an operator $\hat{A}$,
		\begin{equation}
			\hat{A}|a_i\rangle = a_i |a_i\rangle \ ,
		\end{equation}
		where we have labeled the eigenstates by their corresponding eigenvalue. A device that evolves each eigenstate $|a_i\rangle$ into a unique position $|x_i\rangle$ with probability 1 is said to \emph{measure} the operator \cite{caticha:2000}. In fact, the device would be a measurement of any operator of the form
		\begin{equation}
			\hat{A} = \sum_i \lambda_i |a_i\rangle\langle a_i| \ ,
		\end{equation}
		for any choice of $\lambda_i$.
		
		This evolution of the system is described by the unitary operator representing the potential of the measurement device,
		\begin{equation}
			\hat{U}|a_i\rangle = |x_i\rangle \ .
		\end{equation}
		There is no need to specify the evolution beyond this; the unitarity of the evolution guaranteed by the Schr\"odinger equation is sufficient. We see that the measurement device essentially maps eigenvectors of the operator $\hat{A}$ to positions. This correspondence can be symbolized by a one-to-one function, $a_i = g(x_i)$. The exact form of the correspondence function depends on the particular measurement device.

		If the eigenstates of the operator $\hat{A}$ form a complete, orthogonal basis then the identity operator is given by the completeness relation,
		\begin{equation}
			\label{eq:discrete_completeness2}
			\sum_i |a_i\rangle\langle a_i| = I \ ,
		\end{equation}
		which allows us to expand an arbitrary wave function as a superposition of the eigenstates,
		\begin{equation}
			|\Psi\rangle = \sum_i c_i |a_i\rangle \ ,
		\end{equation}
		where $c_i = \langle a_i | \Psi\rangle$ are complex coefficients. We are not assuming the Born rule for the operator $\hat{A}$. At this point, the $c_i$ are merely expansion coefficients and have no probabilistic interpretation.
		
		Now, we apply the unitary operator of the measuring apparatus,
		\begin{equation}
			|\Psi'\rangle = \hat{U}|\Psi\rangle = \sum_i c_i |x_i\rangle \ .
		\end{equation}
		We see that the apparatus has evolved the wave function into a superposition of positions. Multiplying on the left by $\langle x_j|$ gives the probability amplitude,
		\begin{equation}
			\langle x_j|\Psi'\rangle = \sum_i c_i \delta_{ij} = c_j \ ,
		\end{equation}
		which implies that the probability of finding the particle at the position $x_j$ is
		\begin{equation}
			p_j = |\langle x_j|\Psi'\rangle|^2 = |c_j|^2 \ .
		\end{equation}
		The complex coefficients then determine the probability that the particle will be found at the position corresponding to the eigenvector for such a measuring apparatus.
		
		There is an additional twist here. If we note the orthogonality of the eigenstates of the operator $\hat{A}$, we can determine this probability from the original wave function,
		\begin{equation}
			|\langle a_j|\Psi\rangle|^2 = |c_j|^2 = p_j \ ,
		\end{equation}
		which \emph{is} Born's rule for the operator $\hat{A}$. Born's rule is a postulate in the standard interpretation of quantum mechanics, but here we show that it is an inevitable consequence of the unitarity of the evolution. The versatility of this result should not be underestimated. Recall that we did not specify measuring apparatus beyond the unitary action on the eigenstates. This result is valid for \emph{any} apparatus that is said to measure the operator $\hat{A}$. The only difference between measurement devices for the same operator is the correspondence function $a_i = g(x_i)$.
		
		Our interpretation of Born's rule is actually quite limited. It simply states that if you can expand a wave function as a superposition of eigenstates of an operator, then the square modulus of the complex expansion coefficients is equal to the probability of finding the particle at the point corresponding to that particular eigenstate following a measurement of the operator. It does not imply that the wave function was originally in the particular eigenstate corresponding to the final position. This is a common point of confusion in the standard QM approach. The confusion seems to arise from the fact that a \emph{mixture} of pure eigenstates with probabilities equal to the square modulus of the expansion coefficients is indistinguishable from the superposition for that \emph{particular} measurement. However, superpositions are not mixtures. Applying a measurement of a \emph{different} kind would not yield identical results. In a paper discussing this very fact, Jaynes was particularly critical of those who assert the viewpoint that a measurement somehow uncovers some physical reality of the measured wave function \cite{jaynes:1980}. He writes:
		\begin{quote}
			It is pretty clear why present quantum theory not only does not use---it does not even dare to mention---the notion of a `real physical situation.' Defenders of the theory say that this notion is philosophically naive, a throwback to outmoded ways of thinking, and that recognition of this constitutes deep new wisdom about the nature of human knowledge. I say that it constitutes a violent irrationality, that somewhere in this theory the distinction between reality and our knowledge of reality has become lost, and the result has more the character of medieval necromancy than of science.
		\end{quote}
		We agree with Jaynes' criticism to an extent. The distinction between information and reality is critical, and imparting reality to the wave function is a risky endeavor. However, in entropic dynamics we do assert that position \emph{is} the sole observable.
		
		Another common misunderstanding of measurement is that it leaves the system in the eigenstate that it was measured in, which is the source of the projection postulate. In actuality, it is an extremely special case known as \emph{filtering}, which will be discussed later in this chapter.

		We will now extend our result to operators with continuous eigenvalues. Consider an operator $\hat{A}$ with a continuous spectrum of eigenvalues,
		\begin{equation}
			\hat{A}|a\rangle = a|a\rangle \ .
		\end{equation}
		We will start by considering discrete $a_i$ and $x_i$ and take the limit as the spacing between points goes to 0. We can write the completeness relation (\ref{eq:discrete_completeness2}) as
		\begin{equation}
			\sum_i \Delta a\ \frac{|a_i\rangle}{\Delta a^{1/2}} \frac{\langle a_i|}{\Delta a^{1/2}} = I \ ,
		\end{equation}
		so that when $\Delta a = a_{i+1} - a_i \rightarrow 0$, this becomes
		\begin{equation}
			\label{eq:continuous_completeness}
			\int\!da\ |a\rangle \langle a| = I
		\end{equation}
		given
		\begin{equation}
			\frac{|a_i\rangle}{\Delta a^{1/2}}\ \rightarrow \ |a\rangle  \ .
		\end{equation}
		Here $\Delta a^{1/2} = (\Delta a)^{1/2}$ and not $\Delta (a^{1/2})$.

		We again consider a measurement device that evolves eigenstates of $\hat{A}$ into unique positions,
		\begin{equation}
			U|a\rangle = |x\rangle \ .
		\end{equation}
		In the continuum, however, the correspondence function $a = g(x)$ must be monotonic and not simply one-to-one in order for the probability densities to be well-behaved. In the limit $\Delta x \rightarrow 0$, the orthogonality of position states is expressed by a Dirac delta distribution,
		\begin{equation}
			\frac{\langle x_i|}{\Delta x^{1/2}}\frac{|x_j\rangle}{\Delta x^{1/2}} = \frac{\delta_{ij}}{\Delta x} \quad \rightarrow \quad \langle x|x'\rangle = \delta(x-x') \ .
		\end{equation}

		We can use the completeness relation to expand an arbitrary wave function as a superposition of the eigenstates of $\hat{A}$,
		\begin{equation}
			|\Psi\rangle = \sum_i \Delta a\ \frac{|a_i\rangle}{\Delta a^{1/2}} \frac{\langle a_i|\Psi\rangle}{\Delta a^{1/2}} \ ,
		\end{equation}
		where the expansion coefficients,
		\begin{equation}
			\frac{\langle a_i|\Psi\rangle}{\Delta a^{1/2}} \quad \rightarrow \quad \langle a|\Psi\rangle = c(a) \ .
		\end{equation}
		Applying the unitary evolution of the wave function gives
		\begin{equation}
			|\Psi'\rangle = \sum_i \Delta a\ \frac{|x_i\rangle}{\Delta a^{1/2}} \frac{\langle a_i|\Psi\rangle}{\Delta a^{1/2}} \ .
		\end{equation}
		We want to know the probability that the system will evolve to some position $x_j$. Multiplying on the left by $\langle x_j|$ and dividing by $\Delta x^{1/2}$ so that we recover the proper limit yields
		\begin{equation}
			\frac{\langle x_j|\Psi'\rangle}{\Delta x^{1/2}} = \sum_i \Delta a\ \frac{\langle x_j|x_i\rangle}{\Delta a^{1/2} \Delta x^{1/2}} \frac{\langle a_i|\Psi\rangle}{\Delta a^{1/2}} \ .
		\end{equation}
		As a last step before taking the limits, we rearrange the expression in order to achieve a change of variables,
		\begin{equation}
			\frac{\langle x_j|\Psi'\rangle}{\Delta x^{1/2}} = \sum_i \Delta x\ \frac{\langle x_j|x_i\rangle}{\Delta x} \frac{\langle a_i|\Psi\rangle}{\Delta a^{1/2}} \left( \frac{\Delta a}{\Delta x} \right)^{1/2} \ .
		\end{equation}
		Finally, taking the limit as $\Delta a \rightarrow 0$ and $\Delta x \rightarrow 0$ and recalling that $a = g(x)$ gives
		\begin{equation}
			\langle x|\Psi'\rangle = \int\!dx'\ \delta(x-x')\, c(g(x')) \left( \frac{dg}{dx'} \right)^{1/2} \ .
		\end{equation}
		Eliminating the integration with the delta function and taking the square modulus yields the probability density,
		\begin{equation}
			\rho'(x) = |\langle x|\Psi'\rangle|^2 = |c(g(x))|^2 \left| \frac{dg}{dx} \right| \ .
		\end{equation}
		Once again the complex expansion coefficients $c(a)$ play a central role in the determination of the probabilities of the resulting positions. We can go one step further by making a change of variables using the correspondence function. A change of variables preserves the probability,
		\begin{equation}
			\rho'(x) |dx| = \rho_{\hat{A}}(a) |da| \ ,
		\end{equation}
		which leads to a probability density in terms of the eigenvalues $a$,
		\begin{equation}
			\rho_{\hat{A}}(a) = |c(a)|^2 \ .
		\end{equation}
		This is still a probability density for the position of the particle \emph{after} measurement. We have simply relabeled the positions with their corresponding eigenvalues.

		Just as in the discrete case, the probability density could be determined simply from the original wave function and the orthogonality of the eigenstates,
		\begin{equation}
			\label{eq:born_continuous}
			|\langle a|\Psi\rangle|^2 = |c(a)|^2 = \rho_{\hat{A}}(a) \ ,
		\end{equation}
		which is the continuous form of Born's rule for the operator $\hat{A}$.

	\subsection{Expectation Values}
	
		Once we have a probability density for the position of the particle after measurement, we are free to calculate any expected values of the position $x$ or functions of the position. One particularly interesting expectation value is that of the correspondence function $a = g(x)$,
		\begin{equation}
			\bar{A} = \int\!dx\ g(x)\rho'(x) \ .
		\end{equation}
		We can apply a change of variables using the correspondence function,
		\begin{equation}
			\bar{A} = \int\!da\ a\rho_{\hat{A}}(a) \ ,
		\end{equation}
		which implies that we are calculating the expected value of the eigenvalues of the operator $\hat{A}$. Replacing the probability density using the Born rule (\ref{eq:born_continuous}) gives us
		\begin{equation}
			\bar{A} = \int\!da\ a\langle \Psi|a\rangle\langle a|\Psi\rangle \ .
		\end{equation}
		Using the eigenvalue equation, $\hat{A}|a\rangle = a|a\rangle$,
		\begin{equation}
			\bar{A} = \langle \Psi|\hat{A}\int\!da\ |a\rangle\langle a|\Psi\rangle \ .
		\end{equation}
		Finally, applying the completeness relation (\ref{eq:continuous_completeness}), results in an very convenient, compact expression for the expectation value,
		\begin{equation}
			\bar{A} = \langle \Psi|\hat{A}|\Psi \rangle \equiv \langle \hat{A}\rangle \ .
		\end{equation}

		One should be careful in the interpretation of this expectation value. It is the expected value of a function of the position of the particle after the measurement. If, however, our correspondence function is linear then
		\begin{equation}
			\langle \hat{A}\rangle = g(\langle x \rangle) \ .
		\end{equation}
		This result is quite remarkable. It states that if there is a linear correspondence between eigenvalues of an operator and resulting positions for a measurement apparatus, the expected value of the resulting position is simply given by the expected value of the operator, provided the coordinates are properly transformed. We are always free to choose a measurement device that satisfies this linear correspondence.

	\subsection{Amplification}
		\label{subsec:measure_amp}
	
		Now that we have cast all measurements as position measurements, we are left with an engineering problem. How do we determine the position of a microscopic particle undergoing Brownian motion? The solution is to somehow couple the microscopic system of interests to a macroscopic \emph{amplification} system. The classical nature of the amplifier will be discussed later in this chapter when we discuss the classical limit.
		
		An amplification system is generally set up in an initial unstable equilibrium. When the position of system of interest activates the amplifying system, there is a cascade reaction that leaves the amplifier in a macroscopically distinguishable final state. For example, a photomultiplier tube detects the presence of a single photon with just such a cascading effect. An incident photon ejects electrons from a photoelectric material. The electrons are then accelerated into an electrode, ejecting even more electrons. The process repeats many times so that small but measurable current is detected.
		
		By design, the amplification process does not interfere with the result of the measurement. That is, if an eigenstate will evolve to a position $x_r$ without the amplifier's presence, then it will evolve to $x_r$ when it is present. The measurement that maps the eigenstates of an operator to positions is complete before the amplification process takes over. Therefore, if the goal of the inference is merely to determine the position of the particle, then it is not appropriate to form a superposition of the macroscopic amplifier and the system of interest as in (\ref{eq:meas_prob_superpos}).
		
		Mathematically, we are concerned with the joint probability $P(x_r, \alpha_r)$ of the position of the particle $x_r$ after measurement and the subsequent macroscopic state of the amplifier $\alpha_r$ that corresponds to that position. The probability that the amplifier will be in the state $\alpha_r$ is given by simple rules for probabilities,
		\begin{equation}
			P(\alpha_r) = \frac{P(x_r)P(\alpha_r|x_r)}{P(x_r|\alpha_r)} \ .
		\end{equation}
		The probability of the particle's position $P(x_r)$ is given by Born's rule. The marginal probabilities are \emph{designed} to be as close to 1 as possible, $P(\alpha_r|x_r) \approx P(x_r|\alpha_r) \approx 1$, so that $P(\alpha_r) \approx P(x_r) = |\langle a_r|\Psi \rangle|^2$. This is the requirement for a \emph{good} amplification device.

		It may seem that we are simply redrawing von Neumann's line between the classical and the quantum with our treatment of the amplifying system. In some sense, we are doing just that. However, the line here is not between a classical ``reality'' and a quantum ``reality''---it is between the microscopic system of interest and the amplifying system, whose microscopic degrees of freedom are of no interest. These are informationally different situations. If the wave function represents a `real physical situation,' there is no justification for such a treatment. Entropic dynamics, however, is operating at the epistemological level. Not only are we free to change the relevant information when the question changes, we are compelled to.
		
		In the quantum measurement part of the procedure, the relevant information demands that we keep track of the microscopic degrees of freedom in order to make proper inferences about the position of the system of interest. In the amplification portion of the procedure, the question has changed. The microscopic details of the amplifier are not relevant information, which allows us to treat the amplifier classically. Perhaps this is why von Neumann's line was such a successful approach.
		
		While we assert that the microscopic details of the amplification apparatus does not constitute relevant information, there is, in fact, no reason why we could not treat the amplifier as a quantum system as well; it is simply not necessary. Consider the entangled system consisting of a particle at $x_r$ \emph{after} a measurement and an amplifying apparatus in an initial state,
		\begin{equation}
			|\Phi\rangle = |x_r\rangle |0\rangle \ .
		\end{equation}
		The apparatus could be in the superposition,
		\begin{equation}
			|0\rangle = \int\!dm\ c(m) |0,m\rangle \ ,
		\end{equation}
		where $m$ indicates one of the multitude of configurations of the amplifier that is consistent with being in its initial state. Consider the unitary evolution that evolves the entangled system,
		\begin{equation}
			\hat{U}|x_r\rangle|0,m\rangle = |X_{r,m}\rangle \ ,
		\end{equation}
		where the position $X_{r,m}$ is a position in the joint configuration space of both the particle of interest and the amplifier. Here $r$ indicates a region of the configuration space that is consistent with the amplifier being in the macroscopic state $\alpha_r$.

		The full evolution of the entangled system is then
		\begin{equation}
			\hat{U}|\Phi\rangle = \int\!dm\ c(m) |X_{r,m}\rangle \ ,
		\end{equation}
		where $c(m)$ are the expansion coefficients of the initial state of the amplifier. This implies that the probability of finding the entangled system at a position $X_{r,m'}$ given that the particle was at $x_r$ is
		\begin{equation}
			P(X_{r,m'}|x_r) = |c(m')|^2\, dm' \ .
		\end{equation}
		However, we neither can nor desire to know whether the system is in a particular configuration specified by $m'$. So we marginalize over the irrelevant degrees of freedom $m'$. Integrating gives
		\begin{equation}
			P(\alpha_r|x_r) = \int\!dm'\ |c(m')|^2 = 1 \ ,
		\end{equation}
		the probability of finding the particle in the macroscopic pointer state $\alpha_r$ given that the particle was measured at $x_r$. We see that there is no reason why we could not treat the amplifier as a quantum system, but since the microscopic degrees of freedom are not relevant, there is no need.
	
	\subsection{Filtering}

		A special type of experimental procedure is called \emph{filtering}. Consider a measurement of an operator $\hat{A}$. The measurement device evolves the wave function to positions,
		\begin{equation}
			\hat{U}|\psi\rangle = \sum_i c_i \hat{U}|a_i\rangle = \sum_i c_i |x_i\rangle \ .
		\end{equation}%
		\begin{figure}
			\centering
			\includegraphics[scale=0.75]{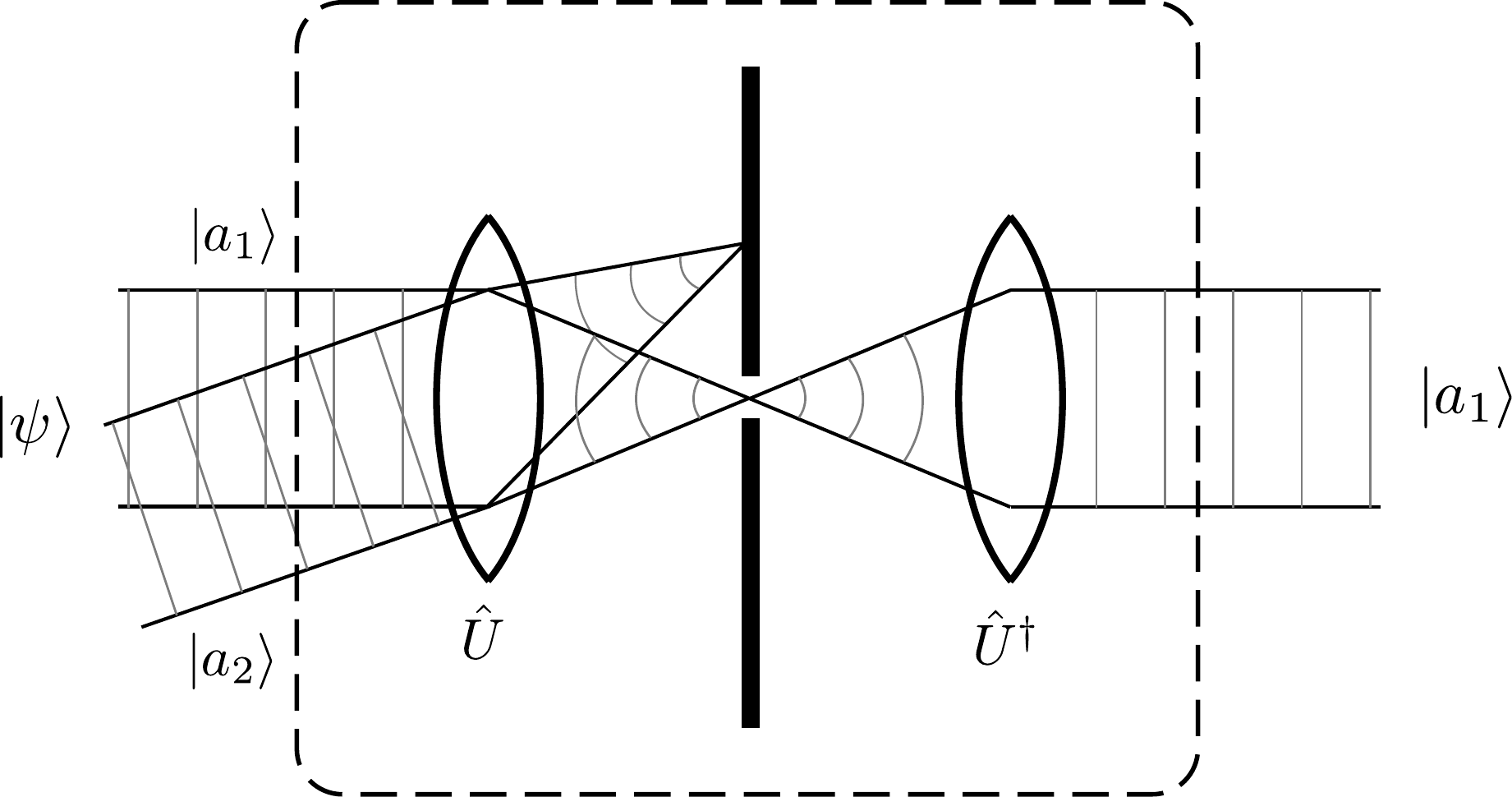}
			\caption[Depiction of a filtering procedure]{\label{fig:filter}A depiction of a filtering procedure. A measurement device is depicted as a `lens' that focuses eigenstates to unique points with the evolution operator $\hat{U}$. The wave function $|\psi\rangle$ is a superposition of two eigenstates. After filtering out one of the final position states, applying inverse evolution with $\hat{U}^\dagger$ evolves the wave function to an eigenstate.}
		\end{figure}%
		Now we will use a screen to block all but perhaps one position. Then we apply an inverse operation that further evolves the particle exiting the hole in the screen,
		\begin{equation}
			\hat{U}^\dagger |x_i\rangle = |a_i\rangle \ ,
		\end{equation}
		so that the wave function exiting the complete apparatus is one of the original eigenstates. This filtering procedure is depicted in figure \ref{fig:filter}.
		If the wave function exiting a filter is subjected to a subsequent measurement of the operator $\hat{A}$, it will naturally evolve to a single position $x_i$ with probability 1. This is the source of the repeated measurement condition that leads one to the projection postulate. 
		
		What is the justification for collapsing the wave function from a superposition of position states to a single position corresponding to the hole in the screen? The answer is very straightforward. If we wish to discuss the dynamics of a particle exiting the filtering device, we need to include the highly relevant information that the particle must have originated from the point $x_i$. The subsequent inverse unitary evolution demands that the particle be in the eigenstate $|a_i\rangle$ upon leaving the apparatus. In fact, there is no reason why we could not consider the filter as a part of the preparation procedure. Then we would simply say that the relevant information is that the wave function satisfies the eigenvalue equation for that operator, $\hat{A}|a_i\rangle = a_i|a_i\rangle$, which amounts to fully constrained updating with the ME method. Updating the wave function in this way in only possible in entropic dynamics. In EQD, the phase is constructed purely from probabilities. If additional information is available, then the probabilities and, in turn, the phase update.
		
	\subsection{Updating the Wave Function}
		
		Fully-constrained updating is not the only means of updating a wave function. Since the wave function codifies probabilistic information about the $x$ and $y$ variables, we can update the wave function when we learn any kind of information about these variables. For example, consider the situation where the only relevant information is that, after interacting with a filter, the probability density is some known function $\rho'(x) = \rho_D(x)$. We learn no information about the extra $y$ variables. How should the wave function update? 
		
		We seek to find the joint posterior $P(x,y) = \rho'(x)p'(y|x)$ that maximizes the joint entropy,
		\begin{equation}
			\mathcal{S}[P,Q] = -\int\!dx\,dy\ P(x,y) \log \frac{P(x,y)}{Q(x,y)} \ ,
		\end{equation}
		subject to normalization and the constraint on the posterior position density,
		\begin{equation}
			\label{eq:position_constraint}
			\rho'(x) = \int\!dy\ P(x,y) = \rho_D(x) \ .
		\end{equation}
		The posterior that maximizes the entropy is
		\begin{equation}
			P(x,y) = \frac{1}{Z} Q(x,y) e^{-\lambda(x)} \ ,
		\end{equation}
		where $Z$ is a normalization factor and $\lambda(x)$ is a Lagrange multiplier for the position constraint. If we substitute our posterior into the constraint on the position density (\ref{eq:position_constraint}) we find
		\begin{equation}
			\frac{1}{Z}e^{-\lambda(x)} = \frac{\rho_D(x)}{\rho(x)} \ .
		\end{equation}
		Substituting this into the posterior gives
		\begin{equation}
			P(x,y) = Q(x,y) \frac{\rho_D(x)}{\rho(x)} \ .
		\end{equation}
		Expanding the joint probabilities with the product rule gives
		\begin{equation}
			\rho_D(x)p'(y|x) = \rho(x)p(y|x)\frac{\rho_D(x)}{\rho(x)} \ ,
		\end{equation}
		which implies that the $y$ variable distribution does not update, $p'(y|x) = p(y|x)$. Furthermore, if the $y$ distribution does not change, neither does the entropy field. Therefore, our updated wave function is
		\begin{equation}
			\Psi' = \rho_D e^{i\phi'} \ ,
		\end{equation}
		with the updated phase
		\begin{equation}
			\phi' = \phi - \frac{1}{2}\log \frac{\rho_D}{\rho} \ .
		\end{equation}
		It is not yet clear how a filter that behaves in this way would be constructed. However, if a filter does indeed behave in this way, this is how one would treat it.

\section{Classical Determinism}
	\label{sec:measure_classical}
	We saw in section \ref{subsec:measure_amp} that classical systems play an important role in the measurement process: they are used as amplifiers to read out the position of a quantum system. In this section we discuss how the apparently deterministic macroscopic world arises from a quantum system undergoing Brownian fluctuations.
	
	The determinism of macroscopic degrees of freedom of a classical system has been noted numerous times before. Consider a system of $N$ particles. The dynamics of the system occur in $3N$-dimensional configuration space. However, we can write the Schr\"odinger equation for the system in terms of the center of mass coordinates \cite{allori:2009}. For $i \in (1,2,3)$,
	\begin{equation}
		\label{eq:center_of_mass}
		R^i = \frac{1}{N} \sum_{n=1}^N \frac{m_n}{\bar{m}} x^{i_n} \ , 
	\end{equation}
	where $x^{i_n}$ is the $i$th coordinate of the $n$th particle. The average mass $\bar{m}$ is given by $\sum_n m_n/N$. In center of mass coordinates, the Schr\"odinger equation can be written as
	\begin{equation}
		i\hbar\Psi + \frac{\hbar^2}{2M} \nabla^2\Psi + \bar{V} = 0 \ ,
	\end{equation}
	where $M = N\bar{m}$ and $\bar{V}$ is the average potential energy. We can write this as a phase equation
	\begin{equation}
		\hbar \dot{\phi} + \frac{\hbar^2}{2M}(\partial_i\phi)^2 + \bar{V} - \frac{\hbar^2}{2M} \frac{\nabla^2 \rho^{1/2}}{\rho^{1/2}} = 0 \ ,
	\end{equation}
	with corresponding Fokker-Planck equation,
	\begin{equation}
		\dot{\rho} = -\partial^i \left( \rho \frac{\hbar}{M}\partial_i\phi \right) \ .
	\end{equation}
	We see that the center of mass of the system behaves like a particle with a very large mass $M$. From the Fokker-Plank equation, we see that for motion to occur $\dot{\rho} \ne 0$. Therefore, as $M$ gets increases, $\hbar\partial_i\phi$ must increase accordingly. If we write $S_{\text{HJ}} = \hbar \phi$ and take the limit as $M$ gets large, the quantum potential term in the phase equation vanishes. This reduces the phase equation to the Hamilton-Jacobi equation,
	\begin{equation}
		\dot{S}_{\text{HJ}} + \frac{1}{2M}(\partial_iS_{\text{HJ}})^2 + \bar{V} = 0 \ ,
	\end{equation}
	which is an equation of \emph{classical} motion \cite{caticha:2010}.
	
	This classical limit of a system with many degrees of freedom is not new. However, in entropic dynamics we can see it enter at an extremely early point in the development of the theory. Recall the probability that a system will transition from a position $x$ to a position $x'$ in the $3N$-dimensional configuration space (\ref{eq:approx_trans_prob}). We wish to write the transition probability in terms of the center of mass coordinates (\ref{eq:center_of_mass}),
	\begin{equation}
		P(R'|R) = \int\!d^{3N}\!\!x\ P(x'|x)\, \delta\left(\Delta \vec{R} - \Delta \vec{\bar{R}} - \frac{1}{N}\sum_{n=1}^N \frac{m_n}{\bar{m}} (\Delta \vec{x}_n - \Delta \vec{\bar{x}}_n)\right) \ ,
	\end{equation}
	where $\Delta R^i = R'^i - R^i$ is a center of mass displacement, and $\Delta \bar{R}^i$ is the expected center of mass displacement.
	The integration can easily be evaluated using the central limit theorem \cite{caticha:2008}. The resulting center of mass transition probability is
	\begin{equation}
		P(R'|R) = \left( \frac{M}{2\pi\hbar \Delta t} \right)^{3/2} \exp \left[ -\frac{M}{2\hbar\Delta t} \delta_{ij} (\Delta R^i - \Delta \bar{R}^i)(\Delta R^j - \Delta \bar{R}^j) \right] \ .
	\end{equation}
	Again we see that the center of mass behaves like a heavy particle with mass $M$. The expected value of a center of mass step is
	\begin{equation}
		\langle \Delta R^i \rangle = \Delta \bar{R}^i = \frac{\hbar \Delta t}{\bar{m}} \frac{1}{N} \sum_{n=1}^N \partial^{i_n} S(x) \ ,
	\end{equation}
	with fluctuations
	\begin{equation}
		\langle \Delta W^i \Delta W^j\rangle = \frac{\hbar \Delta t}{\bar{m}} \frac{1}{N} \delta^{ij} \ .
	\end{equation}
	The result is the same in any central limit type of problem. While the steps are of order $N^0$ (the $N$ terms in the sum offset the $1/N$), the fluctuations are of order $1/\sqrt{N}$. For large $N$, the fluctuations tend to 0 and the trajectory becomes classical.
	
	The implication of this classical limit is that as while the microscopic degrees of freedom of the system are fluctuating wildly, the macroscopic degrees are deterministic. Classical systems are stable purely by virtue of having many constituent particles.

\section{Momentum and the Uncertainty Principle}

	We now turn our attention to a rather famous result of quantum theory: the uncertainty principle. An in-depth analysis of the uncertainty principle has been performed by Shahid Nawaz \cite{nawaz:2011_1,nawaz:2011_2}. We are only concerned with the implications of the uncertainty principle for measurement.

	The momentum operator in quantum mechanics is defined as $\hat{p}_a = -i\hbar \partial_a$. Consider the expectation value of the operator for an arbitrary wave function $\Psi = \rho^{1/2}e^{i\phi}$,
	\begin{equation}
		\langle \Psi |\hat{p}_a|\Psi \rangle = \hbar\int\!dx\ \rho\, \partial_a \phi - \frac{i\hbar}{2}\int\!dx\ \partial_a \rho \ .
	\end{equation}
	The first integral is an expected value of the gradient of the phase, and the second integral vanishes. Recalling that the current velocity is $v^a = m\partial^a \phi /\hbar$, the expected value of the momentum operator can be written as
	\begin{equation}
		\langle \Psi |\hat{p}_a|\Psi \rangle = m\langle v_a \rangle \ .
	\end{equation}
	We see that the expected value of the momentum operator coincides with the expected value of the \emph{entropic momentum}, $mv^a$.
	
	Now consider the variance of the momentum operator. The variance of some operator $\hat{A}$ is defined as
	\begin{equation}
		\Delta \hat{A} = \langle \hat{A}^2 \rangle - \langle \hat{A} \rangle^2 \ .
	\end{equation}
	The expected value of $\hat{p}^2$ is
	\begin{equation}
		\langle\Psi | \hat{p}^2 |\Psi\rangle = m^2\langle v^2 \rangle + m^2 \langle u^2 \rangle \ ,
	\end{equation}
	so that the variance of the momentum operator is simply
	\begin{equation}
		\label{eq:momentum_operator}
		\Delta \hat{p} = \Delta (mv) + \Delta (mu) \ ,
	\end{equation}
	where we used $\langle u^a\rangle = 0$. We see that while the expectation values coincide, the variances of the momentum operator and the entropic momentum differ by the variance of the \emph{osmotic momentum}.

	The variances of operators are known to obey the \emph{uncertainty principle}. If we have two operators $\hat{A}$ and $\hat{B}$ that do not necessarily commute, we can define a third operator $\hat{C}$ such that
	\begin{equation}
		[\hat{A},\hat{B}] = i\hat{C} \ .
	\end{equation}
	The uncertainty principle states that there is a lower bound on the product of the variances of the two operators \cite{ballentine:1998},
	\begin{equation}
		\Delta \hat{A}\ \Delta \hat{B} \ge \frac{1}{2} |\langle \hat{C} \rangle| \ .
	\end{equation}
	
	For position and momentum operators, the uncertainty principle implies
	\begin{equation}
		\Delta \hat{x}\ \Delta \hat{p} \ge \frac{\hbar}{2} \ .
	\end{equation}
	Examining the variance of the momentum operator (\ref{eq:momentum_operator}), we see how this comes about. As the variance of the position $\Delta x$ gets smaller, the variance of the osmotic momentum increases. So as $\Delta x \rightarrow 0$, the variance of the momentum operator tends to $\infty$.
	
	The interpretation of the uncertainty principle is the subject of some debate \cite{ballentine:1990}. In the orthodox approach to QM, the uncertainty principle is thought to imply that one cannot simultaneously measure position and momentum of a particle to arbitrary accuracy, implying that the wave function cannot simultaneously possess a well-defined position and momentum. However, this is not true for entropic momentum. There is no reason why we cannot construct a wave function such as
	\begin{equation}
		\Psi(x) = \delta(x-x_0) \exp[ik_ax^a] \ ,
	\end{equation}
	where $k_a$ is a constant, and the $\delta$ is not a true Dirac delta but rather some arbitrarily narrow square integrable function. The variance of the position is 0, and the variance of the entropic momentum is also 0. (We are also free to construct such a wave function in standard QM as well.)
	
	So what is the proper interpretation of the uncertainty principle? It simply states that it is not possible to \emph{prepare} a wave function that would have a statistical dispersion of 0 for two non-commuting operators. (If two operators commute, an eigenstate of one operator must also be an eigenstate of the other.) A measurement focuses an eigenstate of an operator to a single measurement point. For the momentum operator, an eigenstate is a plane wave,
	\begin{equation}
		\phi_k(x,t) = \frac{1}{\sqrt{2\pi\hbar}} \exp\left[ik_a\left(x^a-\frac{\hbar}{2m}k^at\right)\right] \ .
	\end{equation}
	The corresponding probability density is uniform. A measurement of position is simply a determination of where the particle is at that instant without interacting with a measurement device. Such a measurement would have infinite variance for a plane wave.
	
	From a maximum entropy method point of view, a state preparation that is simultaneously an eigenstate of two non-commuting operators is an overconstrained system. Each eigenvalue constraint is fully-constraining. That is, given each eigenvalue problem, there is only one possible wave function that satisfies it. If the operators do not commute, these constraints do not lead to the same wave function. Therefore, it is not possible to assign a wave function given the incompatible information.

\section{Sequential Measurements}
	\label{sec:measurement_sequential}
	We finish this chapter with a comment on sequential measurements. If a system is subjected to a filter and then after some time a further measurement of the same type, the correlations between the measurements are called multitime correlations. In standard QM theory, these correlations are simply determined by `collapsing' the wave function to an eigenstate after the filter, allowing it to evolve for a time, then performing a measurement.
	
	It was noted by Grabert et al. that the theory of stochastic mechanics does not yield the same predictions for multitime correlations that standard quantum theory produces \cite{grabert:1979}. This disagreement contributed to Nelson's abandonment of stochastic mechanics. A solution was later proposed by Blanchard et al. that essentially postulates a new Wiener process after the filter \cite{blanchard:1986}. While this does result in the correct predictions, the introduction of the new stochastic process after the filter amounts to a stochastic mechanics version of the projection postulate. Such a postulate is undesirable in any theory, but it is particularly problematic in the stochastic approach. In stochastic mechanics, the Weiner process driving the evolution of the wave function is a very real feature. Demanding that it change upon measurement does not seem to be justified in a theory professing to describe reality itself.
	
	In entropic dynamics, the rules for inference leave no room for any such postulates. The wave function after the filter should be updated with the highly relevant information that the system can only be in the particular state defined by the filter. A subsequent measurement would be in agreement with the standard QM result.

\section{Conclusions}

	In this chapter we showed how entropic dynamics can describe a full theory of measurement despite having only one observable: position. Born's rule for position measurements is an automatic consequence of our statistical model. For other types of measurement, however, we had to derive it as we are not justified in introducing any measurement postulates. We showed how a measurement device unitarily evolves eigenstates of an operator to unique positions. A consequence of this unitary evolution was that the probability for the position of the particle after interacting with the measurement apparatus was given by the complex expansion coefficients of the initial wave function, which is Born's rule for other observables. The rule is simply a convenient means of calculating the probability for a particle's position.

	We also showed that the projection postulate in standard QM is only applicable in the extremely special case of filters. In entropic dynamics, we do not need to postulate a collapse of the wave, even in this special case. Since the phase of the wave function is constructed entirely of probabilities, we are able (and required) to update the phase and the position distribution when new relevant information is available. For a filter, this information is that after interacting with the filter, the wave function must be in the eigenstate corresponding to the filter. Furthermore, we showed how information of a different kind can be used to update the wave function according to the ME method.
	
	In section \ref{sec:measure_classical} we showed how the deterministic classical world arises from the fluctuating quantum world. The macroscopic degrees of freedom of a large system of particles evolve deterministically despite the fluctuations. While this classical limit has been noted before in standard QM, it enters at an extremely early point in entropic dynamics.
	
	Then we discussed the uncertainty principle and momentum. The main result was that the uncertainty principle applies to state preparation, not measurement. It is not possible to prepare a wave function that can evolve to a single point for measurement devices described by two non-commuting observables. Such a preparation would require ME updating with incompatible information.
	
	Finally, we discussed sequential measurements. These measurements were problematic for the theory of stochastic mechanics, which did not give the correct results. In entropic dynamics, however, sequential measurements are easily handled.

\mychapter{Conclusions}
	\label{chapter:conclusions}
	The standard theory of quantum mechanics is built on a handful of postulates. One goal of entropic dynamics is to replace these postulates by more fundamental, informational assumptions. We conclude this thesis with a review of a representative list of postulates underlying standard quantum theory. We examine each one and show that they are either unnecessary or are the result of more fundamental informational assumptions.

\section{The Postulates of Quantum Mechanics}

There is no widely accepted list of postulates in the orthodox approach. However, we will consider a reasonable list by Zettili \cite{zettili:2009} as a representative example:

\newtheorem{qpost}{Postulate}

\begin{qpost}
	The state of a system is described by a complex wave function $\Psi$. Any linear superposition of states is also a state.
\end{qpost}
\noindent In section \ref{sec:eqd_schrod} we showed how the wave function is simply a convenient means of combining the probability density and the phase, $\Psi = \rho^{1/2}e^{i\phi}$. The consequence is that the equation of motion becomes the linear Schr\"odinger equation (\ref{eq:schrod}). While the linearity of the Schr\"odinger equation implies the superposition principle, we do not see the need to demand that every superposition must be physically realizable. However, if one could construct such a state, the state would indeed evolve according to the Schr\"odinger equation.

\begin{qpost}
	A wave function evolves in time according to the Schr\"odinger equation.
\end{qpost}
\noindent As we just noted, the Schr\"odinger equation is a convenient rewriting of the differential equations that govern the motion of a quantum system in configuration space. It is the consequence of making inferences given a key set of relevant information owing to more fundamental assumptions.

\begin{qpost}
	For every observable there is a corresponding Hermitian operator. Upon measurement, the only results of the measurement are the eigenvalues of the operator. After measurement of an eigenvalue, the wave function must be in the corresponding eigenstate, which is the projection postulate.
\end{qpost}
\noindent In entropic dynamics, there is only one observable -- position. A measurement of an operator amounts to mapping eigenstates of the operator to unique positions. The unitary evolution of the wave function ensures that basis vectors of a linear Hermitian operator will evolve to unique positions given the appropriate measuring device. Measuring an eigenvalue simply means that one found the particle at the position that an eigenstate would have gone. Expansion of the wave function into basis vectors is simply a convenient means of determining the probability distribution after interacting with a measurement potential.

Additionally, we reject the projection postulate as a general principle as other authors have \cite{ballentine:1970,ballentine:1990}. The repeatable measurement condition is an extremely special case masquerading as a general principle. In EQD, the special case is a filter, which updates the wave function with the ME method.

\begin{qpost}
	The outcome of a measurement is probabilistic, where the probability of finding the wave function in a given state is determined by the Born rule.
\end{qpost}
\noindent In chapter \ref{chapter:measurement} we demonstrated how the Born rule need not be postulated. For positions, the Born rule is an immediate result of the statistical model. For measurements of other operators, the Born rule was shown to be a straightforward consequence of the unitary evolution of a system in a measurement apparatus. It provides a convenient means of calculating the probabilities of position outcomes and their expected values in a device-independent way.

\section{Final Thoughts}

Throughout this work, there has been one consistent theme -- inference based on incomplete information. We began by describing a means of consistently applying and updating probabilities in order to make inferences. We then applied these methods of inference to quantum theory in the form of entropic quantum dynamics.

In chapter \ref{chapter:symmetry} we discussed the meaning of symmetry from an informational perspective. The major result was that we formulated a single condition that any transformation must satisfy in order to qualify as a symmetry.

Then in chapter \ref{chapter:galilean} we applied our symmetry condition to the extended Galilean transformation. We examined the transformation from a very fundamental point in the development of entropic dynamics. There are two main conclusions to draw from this chapter. First, the dynamics of a quantum system are covariant under the extended Galilean transformation. This covariance is true in standard QM and must be true in EQD as well. Second, an effective gravitational potential appears as a result of the transformation. When the transformation is to a uniformly accelerating frame, the dynamics are indistinguishable from a uniform gravitational potential. The indistinguishability is the result of an equivalence of information in the two different physical situations.

In chapter \ref{chapter:measurement} we developed an entropic theory of measurement. We show that, despite having only position as an observable, we can describe a multitude of measurements. Born's rule for position measurements is an automatic consequence of our statistical model. For other types of measurement, however, we derived it as a consequence of the unitary evolution of the wave function. We see that Born's rule is simply a convenient means of calculating the probability for a particle's position after interacting with a measurement device.

We also discussed the special experimental procedure of filtering. In entropic dynamics, we do not need to postulate a collapse of the wave, even in this special case. Since the phase of the wave function is constructed entirely of probabilities, we are able (and required) to update the phase and the position distribution when new, relevant information is available. For a filter, this information is that after interacting with the filter, the wave function must be in the eigenstate corresponding to the filter. Furthermore, we showed how information of a different kind can be used to update the wave function according to the ME method.

In section \ref{sec:measure_classical} we showed how the deterministic classical world arises from the fluctuating quantum world. The macroscopic degrees of freedom of a large system of particles evolve deterministically despite the fluctuations. In EQD, this classical limit appears at an extremely early point in the development of the theory.

Then we discussed the uncertainty principle and momentum. The main result was that the uncertainty principle applies to state preparation, not measurement. It is not possible to prepare a wave function that can evolve to a single point for measurement devices described by two non-commuting observables as such a preparation would require ME updating with incompatible information.

At this beginning of this chapter, we showed that every postulate in the standard approach to quantum theory can be replaced by more fundamental, epistemological assumptions or eliminated as unnecessary. The theory of entropic quantum mechanics holds a great deal of promise as a theory for non-relativistic quantum mechanics. However, the real success here is that of information and inference. The power of these methods to simplify and clarify our physical theories is impressive, to say the least. The real success will be in moving forward to new and more general theories, where we hope the same inference methods will find great utility.

\backmatter

\bibliographystyle{apsrev4-1long}
\bibliography{dissertation}

\end{document}